\documentclass{article}
\usepackage{fullpage}
\usepackage{color}
\usepackage{amsmath}
\usepackage{amssymb}

\usepackage{hyperref}

\usepackage{tikz}
\usetikzlibrary{shapes,arrows}
\tikzstyle{default} = [draw, minimum size = 3em, text width = 4em,
text centered]
\tikzstyle{wide}=[draw, minimum size=3em, text width=7.5em, text
centered] 
\tikzstyle{narrow}=[draw, minimum size=3em, text
width=2em, text centered]
\tikzstyle{bigS}=[draw, minimum width=23em, minimum height=5em]
\tikzstyle{rounded} = [rectangle, draw, fill=blue!20, text width =
7em, text centered, rounded corners, minimum height=4em]
\newcommand{\kex}{\ensuremath \langle k \rangle_{\text{ex}}}
\newcommand{\I}{\ensuremath \langle I \rangle}
\newcommand{\Sa}{\ensuremath \langle S \rangle}
\newcommand{\ave}[1]{\ensuremath \left\langle #1 \right\rangle}

\newcommand{\order}{\ensuremath \mathcal{O}}
\newcommand{\Ro}{\ensuremath R_0}
\newcommand{\diff}[2]{\frac{\mathrm{d}#1}{\mathrm{d}#2}}
\newcommand{\diffm}[3]{\frac{\mathrm{d}^{#1}#2}{\mathrm{d}#3^{#1}}}

\newcommand\T{\rule{0pt}{2.6ex}}
\newcommand\B{\rule[-1.2ex]{0pt}{0pt}}

\title{Epidemic spread in networks: Existing methods and current
  challenges}
\author{Joel C. Miller$^1$\footnote{JCM dedicates this work to the memory of Bob Borrelli, who taught him how to use integrating factors and much more.}~~and Istvan Z. Kiss$^2$\\
{\footnotesize $^1$ School of Mathematical Sciences and Monash Academy for Cross \& Interdisciplinary Mathematics,}\\[-4pt] {\footnotesize Monash University, Melbourne, VIC 3800, Australia}\\
{\footnotesize $^2$ School of Mathematical and Physical Sciences, Department of Mathematics,}\\[-4pt] {\footnotesize University of Sussex, Falmer, Brighton BN1 9QH, UK}}

\begin{document}
\maketitle
\begin{abstract}
We consider the spread of infectious disease through contact networks of Configuration Model type.  We assume that the disease spreads through contacts and infected individuals recover into an immune state.  We discuss a number of existing mathematical models used to investigate this system, and show relations between the underlying assumptions of the models.  In the process we offer simplifications of some of the existing models.  The distinctions between the underlying assumptions are subtle, and in many if not most cases this subtlety is irrelevant.  Indeed, under appropriate conditions the models are equivalent.  We compare the benefits and disadvantages of the different models, and discuss their application to other populations (\emph{e.g.,} clustered networks).  Finally we discuss ongoing challenges for network-based epidemic modeling.
\end{abstract}

\section{Introduction}
Mathematical models of infectious disease spread have played a significant role in improving our understanding of epidemics and developing better intervention strategies~\cite{andersonmay}.  Invariably, these models make simplifying assumptions in order to arrive at tractable equations.

Recently, considerable research has focused on eliminating some of the standard assumptions of mathematical models.  There has been particular emphasis on understanding how contact networks influence disease spread~\cite{diekmann:network,keeling:networkstructure,kenah:networkbased,kretzschmar:concurrent,meyers:contact,meyers:sars,babak:finite,newman:spread,pastor-satorras:scale-free}.  In this paper we focus on research
into two features of contact networks: Partnerships have nonzero
duration and different individuals have different numbers of
partners.  We will focus our attention on ``SIR'' diseases, diseases
in which individuals begin susceptible, become infected through
partnerships with infected individuals, and may eventually recover
into an immune class.

The models we investigate in this paper consider the spread of an
infectious disease through a population whose partnerships are
static.  The populations have ``Configuration Model'' structure:
each individual is assigned a number of stubs (its degree, $k$),
and then finds partners for each stub randomly from the available
stubs~\cite{MolloyReed,newman:structurereview}.  So if $P(k)$ is the probability a random individual $u$ has $k$
partners, the probability that a partner $v$ has $k'$ partners is $k'
P(k')/\ave{K}$ where $\ave{K}$ is the average degree --- there is a
size bias because an individual's partners are selected proportional
to their degrees.  We assume that infected individuals transmit to each partner as an independent Poisson process of rate $\beta$ and recover as an independent Poisson process of rate $\gamma$.

Several competing systems of equations have been developed to capture the population-scale dynamics of disease spread in these populations.  These models have varying levels of complexity and detail.  Each model makes assumptions about the independence of individuals that appear in the form of a closure.  The assumptions are subtly different.  The resulting models are known to be closely related~\cite{house:insights,taylor:ed_pairwise}.  We introduce these models and provide their derivations.   We clarify these relations and show that in appropriate cases 
the models are identical.  We discuss their advantages and disadvantages for different epidemiological questions and present more recent directions of network-based infectious disease modeling.  We end with a discussion of ongoing challenges.  We include an appendix which shows the mathematical relationships between the models by demonstrating how one can derive some models from others.

\section{The existing models}

We study five different models having three different basic approaches.  The three approaches are characterized as ``pairwise'' models, ``effective degree'' models, and ``edge-based compartmental'' models.  In all cases, the disease is assumed to transmit at rate $\beta$ from an infected individual to its partner (which may or may not be susceptible).  If the partner is susceptible, the transmission results in immediate infection.  Infected individuals recover at rate $\gamma$ and become immune.  One notable model to incorporate partnership duration that we do not investigate is the ``renewal equation'' approach of~\cite{diekmann:network}.

We give a brief heuristic explanation of the approaches here.  The ``pairwise models'' observe that the rate susceptible individuals become infected is proportional to the number of partnerships between susceptible and infected individuals.  The main effort of these models is in tracking how the number of such partnerships change in time.   The ``effective degree'' models focus on the number of individuals with a given number of ``effective'' partnerships.  These models ``discard'' partnerships once it is clear that they will no longer play a role in transmission (for example, if a partner $v$ of $u$ recovers, we no longer have to track the $v$-$u$ partnership).  Consequently these models stratify individuals by their ``effective'' degree, and track the probability an effective partner is infected.  Finally, the ``edge-based compartmental'' models focus on the probability that a partner $v$ has transmitted to $u$, creating compartments showing the probability $v$ has transmitted, or --- if it has not yet transmitted --- whether it is susceptible, infected, or recovered.

\subsection{Closures}
The concept of a ``closure'' comes up across many branches of applied mathematics~\cite{vanderHoef:closure,kolmogorov:1941dissipation,kolmogorov:1941local,richardson:closure,rogers:maximum,sagaut:closure}.  Often we are interested in how some physical quantity is distributed at one scale, but calculating that requires knowledge about its distribution at a larger (or smaller) scale.  However, to calculate that larger (or smaller) scale, requires its distribution at another yet larger (or smaller) scale.  This leads to an infinite cascade of scales, and an infinite sequence of equations.  To truncate this system, an assumption is made that at some scale the distribution can be calculated in terms of the distribution at some previous scale, often by assuming the quantity of interest is randomly distributed at the level of truncation.  This results in a finite system of equations which can be solved.  The error, if any, introduced by the closure determines the accuracy of the solution.  The closures we will use can be thought of as similar to the concept of maximal entropy~\cite{jaynes:maxent}.  We will assume that at some scale in the network there is no useful information contained in larger scales, so we will express our equations in terms of the smaller scale.

For the problem of epidemic spread on networks, our ultimate goal is equations giving the  proportion of the population which is susceptible, infected, or recovered.  An epidemic is an inherently stochastic process, so when we write down deterministic equations, we are implicitly assuming that the actual proportion in each state is closely approximated by the expected number in each state.  Thus our equations are appropriate only in large populations with a sufficiently large number of infections at  the initial time.  If the number of infections is too small, we can typically wait until the number infected has grown sufficiently, and then the equations will be accurate.

The rate at which new infections occur clearly depends on the number of partnerships between susceptible and infected individuals.  In turn, changes in the number of different types of pairs depends on the triples (\emph{e.g.}, an $S$--$S$ pair becomes an $S$-$I$ pair through infection introduced from a third individual).  The frequency of various triples depends on still larger structures.  This cascade of scales suggests a closure is needed.  The various models that have been proposed differ in their choice of closure.  There are three distinct closures which we will consider.  These closures are faithful to the dynamics of the epidemic in the sense that if the initial distribution of infection satisfies the closure  assumptions, then at later time the distribution of infection will still satisfy the closure  assumptions.  This faithfulness is a consequence of the assumption that the contact structure is a Configuration Model network.  So if the contact structure is of a different type, then these closures may fail but other similar closures may apply.

\begin{figure}
\begin{center}
\includegraphics[width=0.4\textwidth]{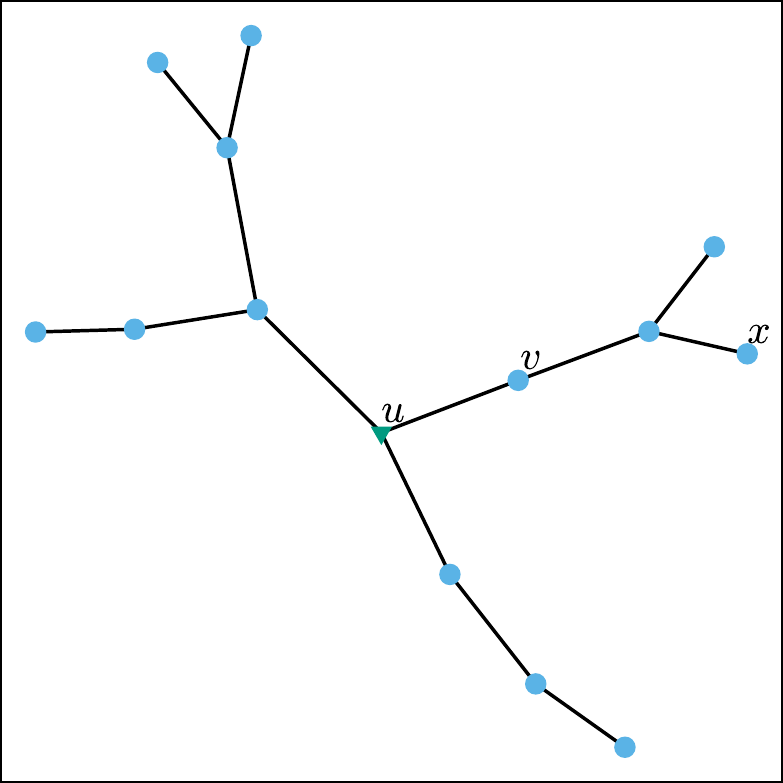}
\end{center}
\caption{\textbf{A demonstration of ``reachable'' individuals}.  Taken from a Configuration Model network with $P(2)=0.8$ and $P(3)=0.2$.  In this case, we choose $u$ and look at individuals within distance $3$ from $u$.  The individual $x$ is reachable from $u$ through $v$, but not through other partners of $u$.  On a sufficiently large scale there will be cycles that connect the different branches.  As the population grows, the size of these cycles grows as well, and the population is locally treelike.}
\label{fig:reachable}
\end{figure}

Although Configuration Model networks have cycles in them, we are interested in the large network limit, for which the networks are ``locally treelike''.  The probability an individual is in a short cycle scales like $1/N$ where $N$ is the population size.  Thus when we look at $u$, we may neglect the distant features and think of the network as a tree with root $u$.  If $v$ is a partner of $u$ and the path from $u$ to $x$ passes through $v$, we refer to $x$ as ``reachable'' from $u$ through $v$, that is $x$ is on the branch of the tree belonging to $v$.  Note that $v$ is considered to be reachable from $u$ through $v$.   This is shown in figure~\ref{fig:reachable}.

We can understand the closures in terms of what information we gain about one partner of a randomly chosen susceptible individual $u$ by considering other partners of $u$.  We now assume that at time $t$, $u$ is a randomly chosen susceptible individual and $v$ is a partner of $u$.  Each of our closures makes a different claim about what information about $v$ (and individuals reachable from $u$ through $v$) we can gain by knowing about other partners of $u$ (and individuals reachable through those partners).  So long as $u$ is susceptible there is no opportunity for dynamics on one branch from $u$ to affect those on another branch.  So we might anticipate that there is no information to be had about $v$ from the other partners of a susceptible individual $u$.  However depending on how the initial infections are chosen, there may be correlations built into the initial condition (examples below).  Depending on what correlations are built into the initial condition, some, or all of our closures may be invalid.  If so, then the equations derived assuming the closure is at best an approximation to the true dynamics.  If there is no correlation built in to the initial condition then all of our closures hold.  Even if there is correlation built in, we anticipate that information about the initial conditions will decay in time, so a closure that does not initially hold is likely to be a good approximation at later time.

 \begin{figure}
\begin{center}
 \includegraphics[width=0.4\textwidth]{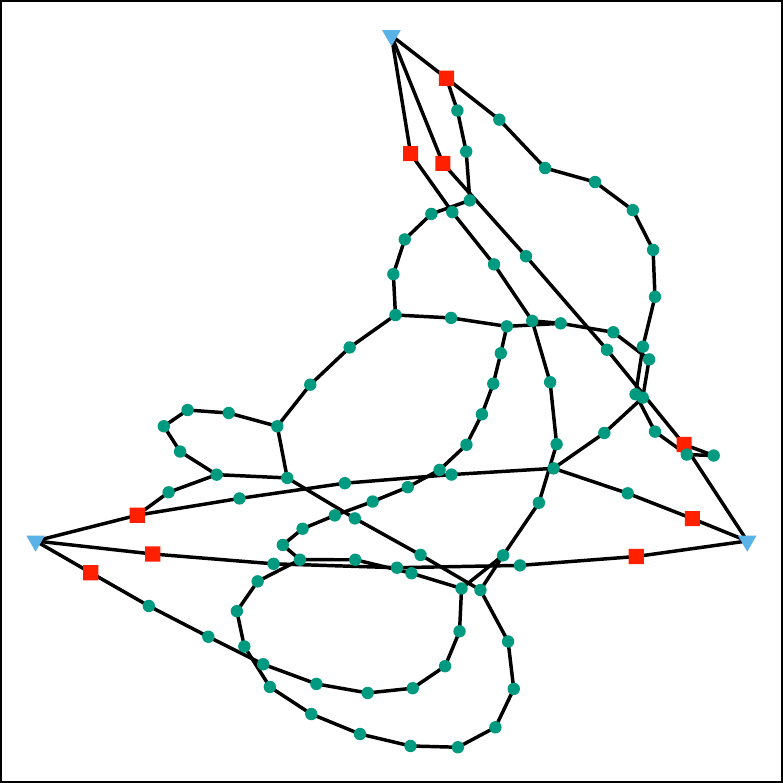}
\qquad
 \includegraphics[width=0.4\textwidth]{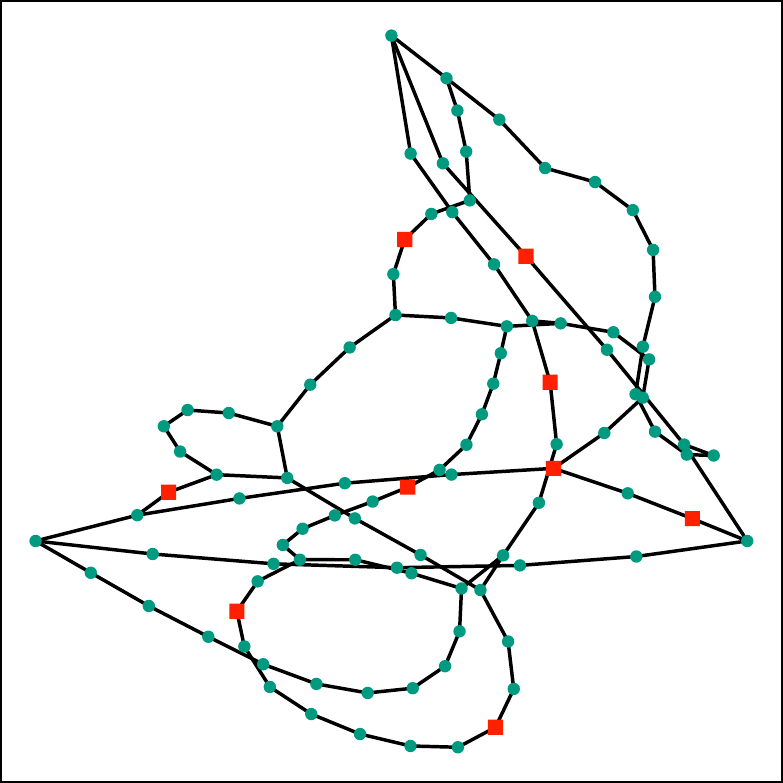}
\end{center}
\caption{\textbf{Two different ways that the initial conditions could be chosen}.  A Configuration Model network is used with $P(2)=08$ and $P(3)=0.2$.  On the left, 3 individuals (denoted by triangles) are chosen uniformly from those with degree three and their partners are infected  (denoted by squares).  On the right, the same number of initial infections are chosen, but randomly from the population.  On the left, we would anticipate slower initial growth for the disease due to ``wasted'' transmissions to the triangular individuals.  The star closure applies for the left, but not the pairs or triples closure.  All closures apply on the right.}
\label{fig:closure_examples}
 \end{figure}

Figure~\ref{fig:closure_examples} demonstrates how a specially selected initial condition can give information about different partners of a random individual.  On the left, $3$ degree three individuals (denoted by triangles in the figure) are chosen from the population and their partners infected.  If we now choose a random individual $u$ from the population and observe whether one partner is infected or not, this gives us information about whether $u$ is one of the ``triangle'' individuals, and this in turn gives information about the probability another partner is infected.  On the right, the same number of initial infections are observed, but they are chosen uniformly from the population.  If we choose an individual $u$ and look at any partner of $u$, it gives no information about the other partners.

\paragraph{Triples closure}
We define a triple to be a set of three individuals for which there is a central individual who is a partner of both of the other two.  Let $[A_pS_kB_q]$ to be the number of triples in the population for which the central individual has status $S$ and degree $k$ and the other two individuals have status $A$ with degree $p$ and status $B$ with degree $q$ ($A$ and $B$ are any of $S$, $I$, or $R$).  Further let $[A_pS_k]$ and $[S_kB_q]$ be the number of pairs of the given types.

The triples closure can be stated as follows: At time $t$ we are given a randomly chosen susceptible individual $u$ and its partner $v$.  The probability that $v$ has a given status may depend on the degrees of $u$ and $v$.  However, information we have about individuals reachable from $u$ through other partners of $u$ does not tell us anything about $v$ or other individuals reachable from $u$ through $v$.  

A consequence of this assumption is that if the triples closure holds at time $t$, then at all later times the frequency of various triples (with susceptible central individual) can be expressed in terms of the number of pairs as
\[
[A_pS_kB_q] = \frac{(k-1)[A_pS_k][S_kB_q]}{k[S_k]}
\]
where $[S_k]$ is the number of susceptible individuals of degree $k$.  So if the triples closure holds, everywhere a triple appears in our infinite cascade of equations, we can reduce it to pairs.  Note that the triples closure as presented here assumes that when we focus on an individual $u$ and its partner $v$, we know both of their degrees but nothing about the degrees of other partners of $u$ or $v$.

To summarize, the assumption of the triples closure is that all relevant details of the epidemic can be deduced by just knowing the frequencies of pairs of different types.

An example of when the triples closure does not hold comes when infection is introduced by selecting some individuals and disproportionately infecting their partners.  

\paragraph{Star closure}
We define a star to be a central individual and all of its partners.  Let $x_{s,i,r}$ denote the number of stars having a central susceptible individual and $s$, $i$, and $r$ susceptible, infected, and recovered peripheral individuals respectively.

The star closure can be stated as follows: If at time $t$ we consider a randomly chosen susceptible individual $u$ and its partner $v$, there can be no information gained about $v$ (or those reachable from $u$ through $v$) by knowledge about individuals on other branches that are at least $2$ steps from $u$.  We do allow the possibility of learning information about the branch through $v$ by knowledge about direct partners of $u$.  However, in the particular case that we know that $v$ was susceptible at time $0$, then we cannot gain information about $v$ from $u$ or the other partners of $u$.

So if we have information about the statuses of all partners of $u$ except $v$, this gives us information about the probability $v$ is susceptible, infected, or recovered.  In those cases in which $v$ is susceptible, there is nothing we can learn about partners of $v$ from knowing about partners of $u$.  In those cases in which $v$ is not susceptible, we may learn about partners of $v$, but this will never affect the status of $v$, so that information is irrelevant to the future status of $u$.  

So the star closure allows for the partners of a susceptible individual to initially have dependent statuses, but the statuses of these partners evolve independently.  So the dependence can only exist due to effects from the initial condition.  

To summarize the star closure, all relevant details of the epidemic can be deduced by just knowing the frequencies of stars of different types having susceptible central individuals.

We will see that the star closure applies in some cases where the triples closure fails.  It is important to note that the converse can also occur: the triples closure may apply in cases where the star closure fails.

\paragraph{Pairs closure}
The pairs closure states that conditional on $u$ being susceptible at time $t$, there is nothing we can learn about $u$ or individuals reachable from $u$ through partners other than $v$ which will give any information about $v$.  So if we know that $u$ is susceptible, a random partner $v$ is as likely to be in any given status as a random partner of any other susceptible individual.

To summarize the pair closure, everywhere a pair involving a susceptible individual appears in our infinite cascade of equations, we can express it in terms of a random partner of a susceptible individual.  Note that if the pairs closure holds, so do the other two closures.

One of the models we will investigate uses a slightly stronger assumpton, which we refer to as the ``strong'' pairs closure.  In this case, we make the assumption that even if $u$ is infected, if there has been no transmission in either direction between $u$ and $v$ we can learn nothing about $v$.  This assumption does not play a major role and appears only in the context of this particular model.

\subsubsection{Validity of the closures}
All of our closures are consistent with the dynamics:  Assume at time $t$ we cannot learn any information about a certain property of $v$ from other partners of $u$ (or individuals reachable through them).  If $u$ is still susceptible at a later time $\hat{t}$, there is no way for information to have travelled between $v$ and the other partners of $u$.  There is no way for any correlations to develop.  This statement can be made more mathematically rigorous by looking at an infinite system of equations which describes the epidemic and verifying that if the closure holds at one time, then it remains valid.  

Note that in all closures, so long as individual $u$ is susceptible, if its partner $v$ is also susceptible, nothing we know about other branches from $u$ can give information about the risk of infection to $v$ (including the degree of $v$).

\subsubsection{Examples of initial conditions for various closures}
We now give some specific examples of initial conditions which may or may not violate some of the closures.

Assume that we color some subset of the individuals blue.  If we infect those individuals at distance $2$ from the blue individuals, none of our closures will hold.  To see this, consider a randomly chosen susceptible individual $u$ and a partner $v$.  By observing whether those individuals a distance $2$ from $u$ which are reachable through partners other than $v$ are infected (or not), we gain information about whether $u$ was colored blue, which gives information about partners of $v$ (and thus the risk of infection for $v$).  This violates all of the closure assumptions.

In contrast, assume that as our initial condition at $t=0$, we infect all degree $3$ individuals which are partners of degree $2$ individuals.  There is no information that goes into this initial condition which depends on structures larger than pairs.  So triples and larger structures can be reduced to pairs.  The triples closure holds.  The other closures do not hold.  The star closure fails because if $u$ is susceptible with degree $3$, we immediately know that its partner $v$ does not have degree $2$, while if $u$ is susceptible with degree $4$ then $v$ may have degree $2$.  Thus if $v$ is a susceptible partner of a susceptible individual $u$, there is information we can learn about $u$ that tells us something about $v$'s degree which in this case informs us about $v$'s partners.  In turn we learn about $v$'s future risk of infection.  So we have learned something about the susceptible partner $v$.  Similarly the pairs closure fails.  If $u$ has degree $3$ and is susceptible, then we learn things about its susceptible partners.  Note that this example implies it is possible to satisfy the triples closure without satisfying the star closure.
 
We can also find conditions which violate the triples closure but satisfy the star closure.  We color some subset of the population blue and at $t=0$ infect all (or some uniformly chosen proportion $p$) of those individuals with at least one blue partner.  Knowing the status of one partner of the susceptible individual $u$ gives information about whether $u$ was colored blue, which in turn gives information about the status of other partners of $u$.  This violates the triples closure.  However, while knowing the status of one partner $w$ of $u$ does give information about another partner $v$, we do not gain any information about $v$ if $v$ is initially susceptible.  So the star closure holds.  The pairs closure again fails for this case.

In all of the above examples, the pairs closure is violated.  However, the violation comes because the initial conditions choose to infect individuals based on the location of some other individual who is likely to initially be susceptible.  This seems rather artificial --- it is difficult to conceive of a relevant scenario where this would happen.  The pairs closure is satisfied in many cases, in particular if we initially infect some individuals uniformly chosen at random.   

In general, even if the initial condition does not satisfy a given closure, we anticipate that as the disease spreads we will lose information about the initial condition.   If so, then the closures will become valid.  So in particular, if the initial condition involves infection of a small proportion of the population, we anticipate that the closures will all be valid by the time the epidemic has grown to appreciable size.  

 \subsection{Pairwise models}
Pairwise models are based on ``pair approximation'' methods.  The rate at which susceptible individuals become infected clearly depends on the number of susceptible-infected partnerships in the population.  If we set $[SS]$ to be the expected number of susceptible-susceptible partnerships in the population and $[SI]$ to be the expected number of susceptible-infected partnerships, we see that there is a flux from $[SS]$ to $[SI]$ resulting from susceptible individuals becoming infected from another partnership.  Determining the rate of this flux is a central focus of pair approximation models.  

We begin by considering the basic pairwise model and then investigate a more recent version which reduces the number of equations required.  Before continuing, we make a note about our accounting.  Each partnership may be thought of as ordered.  Thus it makes sense to talk about the ``first'' or ``focal'' individual of a partnership.  In our accounting, each partnership will appear twice.  So if both partners are susceptible, this partnership will contribute $2$ to $[SS]$, while if one is susceptible and the other infected it will contribute $1$ to each of $[SI]$ and $[IS]$.  Of course, we know $[SI]=[IS]$, so we do not need to calculate both, but we should be aware that the contribution to $[SS]$ is doubled.  

\subsubsection{The basic pairwise model}

Our initial goal is to track individuals of degree $k$ as they pass through different infection classes.  We set $[S_k]$, $[I_k]$, and $[R_k]$ to be the expected number of susceptible, infected, and recovered individuals of degree $k$ in the population.  Because recovery occurs at rate $\gamma$, it is straightforward to see that the flux from $[I_k]$ to $[R_k]$ is $\gamma [I_k]$.  The rate of infection follows easily: Each partnership between a susceptible and an infected individual transmits independently at rate $\beta$, so the flux from $[S_k]$ to $[I_k]$ is $\beta [S_k I]$ where $[S_k I]$ is the total number of partnerships between susceptible individuals of degree $k$ and infected individuals.  As $t$ increases, $[S_kI]$ changes, so we need to be able to track its evolution.

This is the reason that the basic pairwise model~\cite{eames:pair} focuses on pairs or partnerships among individuals.  We set $[S_kS_{k'}]$ to be the total number of partnerships between susceptible individuals of degrees $k$ and $k'$.  This is given by the sum over all susceptible individuals of degree $k$ of the number of susceptible partners of degree $k'$.  This definition gives an implicit direction, so that we can think of this as the total number of partnerships from susceptible individuals of degree $k$ to susceptible individuals of degree $k'$.  If $k=k'$, each partnership is counted twice.

We define $[S_kI_{k'}]$ and $[I_kS_{k'}]$ to correspond to the number of partnerships from susceptible degree $k$ individuals to infected degree $k'$ individuals and from infected degree $k$ individuals to susceptible degree $k'$ individuals respectively.  Note that $[S_kI] = \sum_{k'} [S_kI_{k'}]$.  We similarly define $[S_kR_{k'}]$, $[I_kI_{k'}]$, $[R_kS_{k'}]$, $[I_k R_{k'}]$, $[R_kI_{k'}]$, and $[R_kR_{k'}]$.

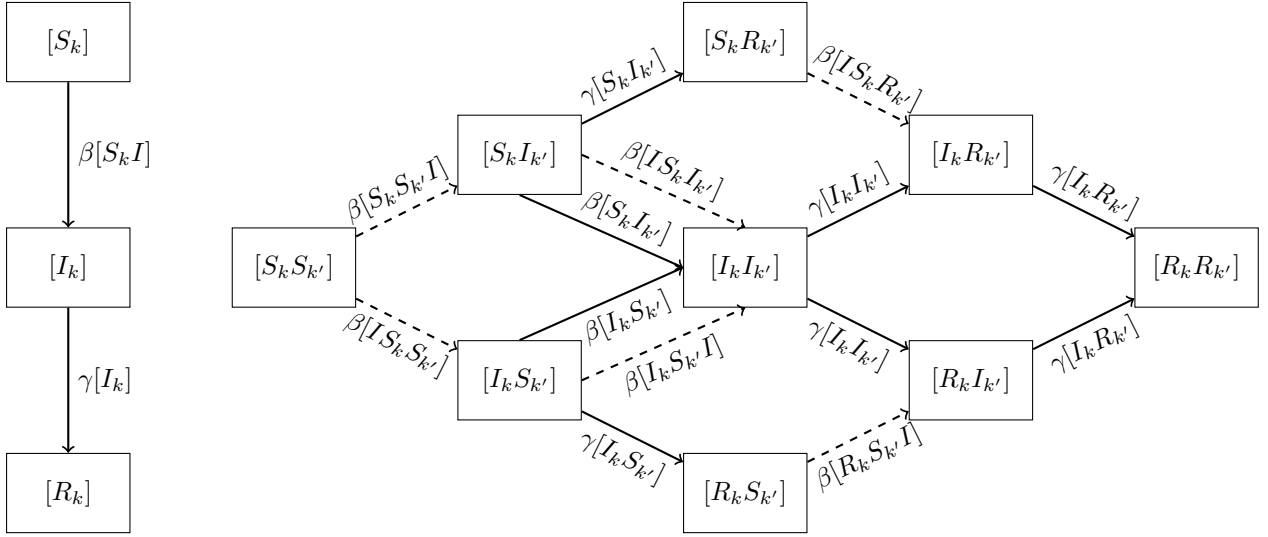
\begin{figure}
\begin{tikzpicture}
\node [default] at (0,-3) (S) {$[S_k]$};
\node [default] at (0,-6) (I) {$[I_k]$};
\node [default] at (0,-9) (R) {$[R_k]$};
\path [->, thick, right] (S) edge node{$\beta [S_kI]$} (I);
\path [->, thick, right] (I) edge node {$\gamma [I_k]$} (R);

\node [default] at (3,-6) (SS) {$[S_kS_{k'}]$};
\node [default] at (6,-4.5) (SI) {$[S_kI_{k'}]$};
\node [default] at (6,-7.5) (IS) {$[I_kS_{k'}]$};
\node [default] at (9,-3) (SR) {$[S_kR_{k'}]$};
\node [default] at (9,-6) (II) {$[I_kI_{k'}]$};
\node [default] at (9,-9) (RS) {$[R_kS_{k'}]$};
\node [default] at (12,-4.5) (IR) {$[I_kR_{k'}]$};
\node [default] at (12,-7.5) (RI) {$[R_kI_{k'}]$};
\node [default] at (15,-6) (RR) {$[R_kR_{k'}]$};
\path [->, thick,above, dashed, sloped] (SS) edge node {$\beta [S_kS_{k'}I]$} (SI);
\path [->, thick,below, dashed, sloped] (SS) edge node {$\beta [IS_kS_{k'}]$} (IS);
\path [->, thick,above, sloped] (SI) edge node {$\gamma [S_kI_{k'}]$} (SR);
\path [->, thick, above, dashed,sloped] (SI.0) edge node {$\beta [IS_kI_{k'}]$} (II.90);
\path [->, thick, above, sloped] (SI.270) edge node[pos=0.6] {$\beta [S_kI_{k'}]$} (II.180);
\path [->, thick, below, sloped] (IS.90) edge node[pos=0.6] {$\beta [I_kS_{k'}]$} (II.180);
\path [->, thick, below, dashed, sloped] (IS.0) edge node {$\beta [I_kS_{k'}I]$} (II.270);
\path [->, thick, below, sloped] (IS) edge node {$\gamma [I_kS_{k'}]$} (RS);
\path [->, thick, above, dashed, sloped] (SR) edge node {$\beta [IS_kR_{k'}]$} (IR);
\path [->, thick, above, sloped] (II) edge node {$\gamma [I_kI_{k'}]$} (IR);
\path [->, thick, below, sloped] (II) edge node {$\gamma [I_kI_{k'}]$} (RI);
\path [->, thick, below, dashed, sloped] (RS) edge node {$\beta [R_kS_{k'}I]$} (RI);
\path [->, thick, above, sloped] (IR) edge node {$\gamma[I_kR_{k'}]$} (RR);
\path [->, thick, below, sloped] (RI) edge node {$\gamma[I_kR_{k'}]$} (RR);
\end{tikzpicture}
\caption{\textbf{The flow diagram underlying the basic pairwise model}.  Dashed
lines denote transitions that rely on infection coming from a source
outside the partnership of interest.  A similar cascade exists at the level of triples, with some transitions depending on paths of length $4$ or on central individuals with three partners.}
\label{fig:PW_basic}
\end{figure}

Putting this together, we get the flow diagrams in figure~\ref{fig:PW_basic}.  We note that when a susceptible individual in a partnership of interest transitions to being infected, it may be because of a transmission from within that partnership or because of a transmission from outside the partnership.  In figure~\ref{fig:PW_basic} we denote transmissions from outside the partnership by dashed lines.  Solid lines are used for transmissions within a partnership or a recovery of an infected individual within the partnership, events that do not depend on triples.

For each dashed line, the central individual of the relevant triple is susceptible.  This means that if the triples closure holds, then we can reduce our triples into pairs.  The resulting equations are then closed at the pairs level:
\begin{align}
[S] &= \sum_k [S_k] \label{eqn:basicalpha}\\
[I] &= \sum_k [I_k]\\
[R] &= \sum_k [R_k] = N-[I]-[S]\\
[IS_k] &= [S_kI] = \sum_{k'} [S_k I_{k'}]\\
\dot{[S_k]} &= - \beta [S_kI]\\
\dot{[I_k]} &= \beta [S_k I] - \gamma [I_k]\\
\dot{[R_k]} &= \gamma [I_k]\\
\dot{[S_kS_{k'}]} &= -\beta \frac{k'-1}{k'}\frac{[S_kS_{k'}][S_{k'}I]}{[S_{k'}]}  - \beta \frac{k-1}{k} \frac{[IS_k][S_kS_{k'}]}{[S_k]} \label{eqn:basicss}\\
\dot{[S_kI_{k'}]} &= \beta
\frac{k'-1}{k'}\frac{[S_kS_{k'}][S_{k'}I]}{[S_{k'}]} - \gamma
[S_kI_{k'}] - \beta\frac{k-1}{k}\frac{[IS_k][S_kI_{k'}]}{[S_k]} -
\beta [S_kI_{k'}]\label{eqn:basicsi}\\
\dot{[S_kR_{k'}]} &= \gamma [S_kI_{k'}] - \beta \frac{k-1}{k}
\frac{[IS_k][S_kR_{k'}]}{[S_k]} \, . \label{eqn:basicomega}
\end{align}
If our goal is to just find $[S]$, $[I]$, and $[R]$, we do not require $[S_kR_{k'}]$, $[I_k]$, or $[R_k]$ directly, so we can avoid calculating
them.  They are included here for completeness.  We can instead write $[I] = N-[S]-[R]$ and $\dot{[R]} = \gamma [I]$.   So we can reduce the system to $2K^2 + K+1$ differential equations where $K$ is the
total number of distinct degrees in the population.  A particular
weakness of these equations is that in many important limiting cases,
$K$ tends to $\infty$.  

It should be noted that if the population structure is not Configuration Model (for example, there may be correlation in the degrees of partners), the triples closure may still hold.  In such cases it may be possible to use a similar process to reduce the model.  Depending on how the population structure deviates from Configuration Model, we may not be able to use the star closure or the pairs closure.  So this approach may be more flexible than the others introduced below, but it requires more equations than some of the later models.

We note that all of these variables are numbers of individuals or numbers of pairs.  Thus they are expected to be proportional to $N$, the population size.  We may divide through by $N$, in which case $S$, $I$, and $R$ become the proportion of individuals in each state.  This may be a more natural system to use as we are interested in the limit of large populations, so by dividing by $N$ we arrive at equations that do not  change with population size.  This is effectively the distinction between intensive and extensive quantities in physical systems.  However, in the existing literature these equations are typically introduced in terms of absolute numbers rather than proportions.

\subsubsection{A compact pairwise model}

A recent paper~\cite{hebert:pathogen} developed some pairwise
equations with a significantly reduced dimension.  It was derived for
a multiple disease model, but we focus on the restriction to just a
single disease.

We begin with the pairwise model using the triples closure, but we
additionally make the pairs closure: We assume that if $v$ is a
partner of a susceptible individual $u$, then the probability $v$ is
susceptible or infected is independent of the degree of $u$ (or other
partners of $u$).

We define $\I$ to be the probability a partner of a susceptible individual $u$ is infected.  By the pairs closure assumption, this is independent of the degree of $u$.  It follows that
\[
\I = \frac{[SI]}{\sum_k k[S_k]} \, .
\]
Since we know that $\sum_k k[S_k] = [SS] + [SI]+[SR]$, this becomes 
\[
 \I = \frac{[SI]}{[SS]+[SI]+[SR]}  \, .
\]

The number of susceptible-susceptible partnerships between degree $k$ and degree $k'$ individuals is equal to the number of partnerships from susceptible individuals of degree $k$ to other susceptible individuals times the probability that other susceptible individual has degree $k'$.  That is
\[
[S_kS_{k'}] = [SS_k] \frac{k' [S_{k'}]}{\sum_{k''}k''[S_{k''}]} \, .
\]
This further leads us to
\[
\frac{\sum_{k'} (k'-1)[S_kS_{k'}] }{[S_kS]} =
\frac{\sum_{k'}k'(k'-1)[S_{k'}]}{\sum_{k''}k''[S_{k''}]} 
\]
which we define to be $\kex$.  This is the average ``excess degree'' of a susceptible individual.  That is, if we reach a susceptible individual by following a partnership this is the expected number of other partners that susceptible individual has.  It appropriately accounts for the fact that the probability of reaching an individual along a partnership is proportional to its degree.  We can simplify the denominator to $[SS]+[SI] + [SR]$.

Using these assumptions the flux $\beta [S_kI]$ from $[S_k]$ to $[I_k]$ can be represented as $\beta k\I[S_k]$.  
\begin{align*}
\dot{[S_k]} &= - \beta k\I [S_k] \\
\dot{[I_k]} &= \beta k \I [I_k] - \gamma [I_k] \, .
\end{align*}
We no longer need $[S_k I]$ explicitly for our
equations at the singles level.  So at the pairs level we can sum 
equations~\eqref{eqn:basicss}--\eqref{eqn:basicomega} over $k$ and $k'$.  We find $\sum_{k,k'}
(k-1) [S_kS_{k'}] [IS_{k}]/k[S_{k}] = \kex \I [SS]$.  Similarly if we switch $k$ and $k'$ in this equation,
we get the same result.  So we arrive at
\begin{align*}
\dot{[SS]} &= -2\beta  \kex \I [SS]\\
\dot{[SI]} &= \beta \kex \I [SS] - (\beta + \gamma) [SI] - \beta \kex \I
[SI]\\
\dot{[SR]} &= \gamma [SI] - \beta \kex \I [SR] \, .
\end{align*}
The $[SR]$ equation can be kept if it simplifies calculations, or we can use $[SR] = (\sum_k k S_k) - [SI]-[SS]$.

\begin{figure}
\begin{tikzpicture}
\node [default] at (0,0) (Sk) {$[S_k]$};
\node [default] at (0,-2) (Ik) {$[I_k]$};
\node [default] at (0,-4) (Rk) {$[R_k]$};
\path [->, thick, right] (Sk) edge node {$k \beta \I [S_k]$} (Ik);
\path [->, thick, right] (Ik) edge node {$\gamma [I_k]$} (Rk);

\node [default] at (6,0) (SS) {$[SS]$};
\node [default] at (6,-2) (SI) {$[SI]$};
\node [default] at (6,-4) (SR) {$[SR]$};
\node [] at (12,0) (SSout) {};
\node [] at (12,-2) (SIout) {};
\node [] at (12,-4) (SRout) {};
\path [->, thick, above] (SS) edge node {$\beta \kex \I  [SS]$} (SSout);
\path [->, thick, right] (SS) edge node {$\beta \kex \I [SS]$} (SI);
\path [->, thick, right] (SI) edge node {$\gamma [SI]$} (SR);
\path [->, thick, above] (SI) edge node {$(\beta \kex \I + \beta)[SI]$} (SIout);
\path [->, thick, above] (SR) edge node {$\beta \kex \I [SR]$} (SRout);
\end{tikzpicture}
\caption{\textbf{The flow diagram for the reduced system
  of~\cite{hebert:pathogen}}.  The $[SS]$, $[SI]$, and $[SR]$
  compartments correspond to the sum of the $[S_kS_{k'}]$,
  $[S_kI_{k'}]$ and $[S_kR_{k'}]$ compartments of the basic pairwise model. The fluxes to the right go to compartments $[IS]$, $[II]$, and $[IR]$ which are not needed for our calculations.}
\label{fig:concise_PW}
\end{figure}
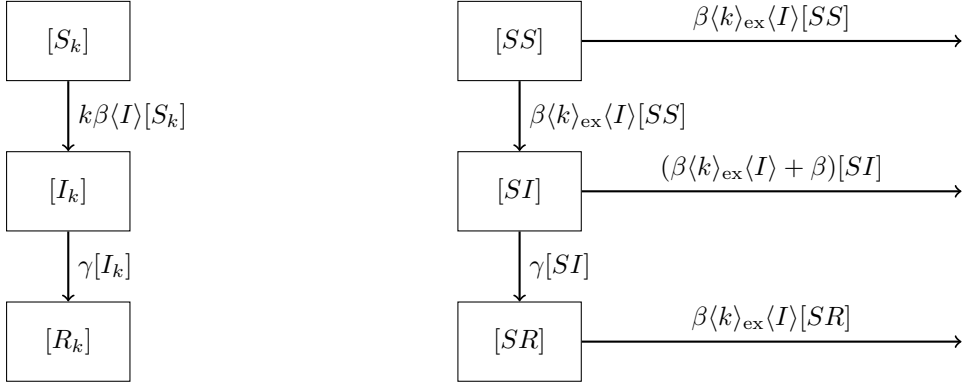

The variables $[I_k]$, $[R_k]$ and $[SR]$ are not needed to find $[S]$, $[I]$, and $[R]$.  If we do not include these, our final system is
\begin{align}
  [S] &= \sum_k [S_k] \label{eqn:compactalpha}\\
  [I] &= N-[S]-[R]\\
  \dot{[R]} &= \gamma [I]\\
  \dot{[S_k]} &= - \beta k\I [S_k] \\
  \dot{[SS]} &= -2\beta  \kex \I [SS]\\
  \dot{[SI]} &= \beta \kex \I [SS] - (\beta + \gamma) [SI] - \beta \kex \I [SI]\\
  \I &= \frac{[SI]}{\sum_k k[S_k]} =
  \frac{[SI]}{[SS] + [SI] + [SR]}\\
  \kex &= \frac{\sum_k k(k-1) [S_k]}{\sum_k k[S_k]} = \frac{\sum_k k(k-1)[S_k]}{[SS]+[SI]+[SR]} \, .\label{eqn:compactomega} \end{align}
The number of differential equations equations needed is reduced to $K+3$.  This model is represented by a reduced flow diagram, shown in figure~\ref{fig:concise_PW}.

Our derivation assumed that the pairs closure holds at all times.  To be fully rigorous, we would additionally need to show that if the pairs closure holds at $t=0$, then the basic pairwise model predicts that it continues to hold at all later time.

Similarly to the basic pairwise model, all of the bracketed quantities
may be divided by the population size $N$ so that the equations are in
terms of ``intensive'' quantities.

\subsection{Effective degree models}

We now turn to ``effective degree'' models.  We define an ``ineffective'' partnership to be a partnership which we know will never transmit infection.  We then stratify individuals by the number of effective partners they have, which we refer to as their effective degree.  How ``omniscient'' we assume we are affects how well we can identify which partnerships will never transmit infection.  Depending on the information we choose to track in our model, we may know that a partner of an infected individual is also infected in which case we know that partnership can be treated as ineffective, or we may simply know that a partner of an infected individual might be infected in which case we cannot discard the partnership and instead must track the probability the partner is infected.  So we can arrive at different models depending on how we choose our variables.

We focus our attention on two models which have been studied already.  By changing the definition of effective partnerships other valid models could be derived.  Using the pairs closure and a careful definition of effective partnerships, we are able to futher simplify these models and introduce a new simpler model.

\subsubsection{The basic effective degree model}
When we introduced the star closure, we used the variable $x_{s,i,r}$ to be the number of susceptible individuals with $s$ susceptible, $i$ infected, and $r$ recovered partners.  The star closure implies that the susceptible partners of one individual are indistinguishable \emph{a priori} from the susceptible partners of another.  In particular the degree of a susceptible individual $u$ gives no information about the risk of infection of its susceptible partners.  

This implies that future risk of an individual $u$ with $s$ susceptible partners and $i$ infected partners is unaffected by the number of recovered partners.  So we can instead use $x_{s,i} = \sum_r x_{s,i,r}$  as our focal variable.  This is the number of stars with a susceptible central individual and $s$ susceptible and $i$ infected peripheral individuals.  We similarly introduce $y_{s,i}$ to correspond to stars with infected central individuals.  These are the variables of~\cite{lindquist}: We assume only a very basic amount of information about which partnerships we can discard, we can only identify those partnerships which contain (at least) one recovered individual.  So effective partnerships are defined to be those partnerships which do not have a recovered member.  The star closure was not defined in terms of infected central individuals, and we shall see that in fact the $y_{s,i}$ variables can be eliminated from these equations.

Here we slightly modify the approach of~\cite{lindquist}: Consider a random effective partnership and choose one individual $u$.  If $u$ is not susceptible repeat until finding a susceptible individual.  Let the partner in this partnership be $v$.  Thus $u$ is chosen from among all susceptible individuals with probability proportional to its effective degree, while $v$ is chosen from all susceptible or infected individuals with probability proportional to its number of susceptible partners.   We define the variables $\xi$ and $\zeta$ to be the expected numbers of infected partners of $u$ given that $v$ is susceptible or infected respectively.  When we eliminate $y_{s,i}$ from the system, $\zeta$ will no longer be needed.  To calculate $\xi$, we must account for the fact that $v$ is chosen with probability proportional to its number of susceptible partners.  Accounting for this bias, we have
\[
\xi =  \frac{\sum_{k_e=1}^M \sum_{s+i=k_e} s i x_{s,i}}{\sum_{k_e=1}^M
 \sum_{s+i=k_e}s x_{s,i}} 
\]
where the summation over $s+i=k_e$ means the sum over all pairs $s$ and $i$ that sum to $k_e$.  We similarly find 
\[
\zeta =  \frac{\sum_{k_e=1}^M \sum_{s+i=k_e} i^2 x_{s,i}}{\sum_{k_e=1}^M
 \sum_{s+i=k_e}i x_{s,i}} \, .
\]
These differ from $G$ and $H$ in~\cite{lindquist} by a factor of $\beta$.  The denominator of $\zeta$ appears different from the denominator of $H$ in~\cite{lindquist} but it is in fact the same\footnote{Here $\sum_{k_e}\sum_{s+i=k_e} ix_{s,i}$ counts the number of susceptible-infected partnerships from the perspective of susceptible individuals, while $\sum_{k_e}\sum_{s+i=k_e} s y_{s,i}$ in~\cite{lindquist} counts the same thing from the perspective of infected individuals}.  Our interpretation of $\zeta$ differs slightly from their interpretation of $H$ because their interpretation requires a slightly stronger assumption than the star closure --- they assume that the star closure applies when the central individual is infected.  This assumption is consistent with the dynamics, but if the initial condition violates this assumption, this assumption will fail at later times.  However, the failure is irrelevant for the purposes of determining the evolution of $S$, $I$, and $R$.

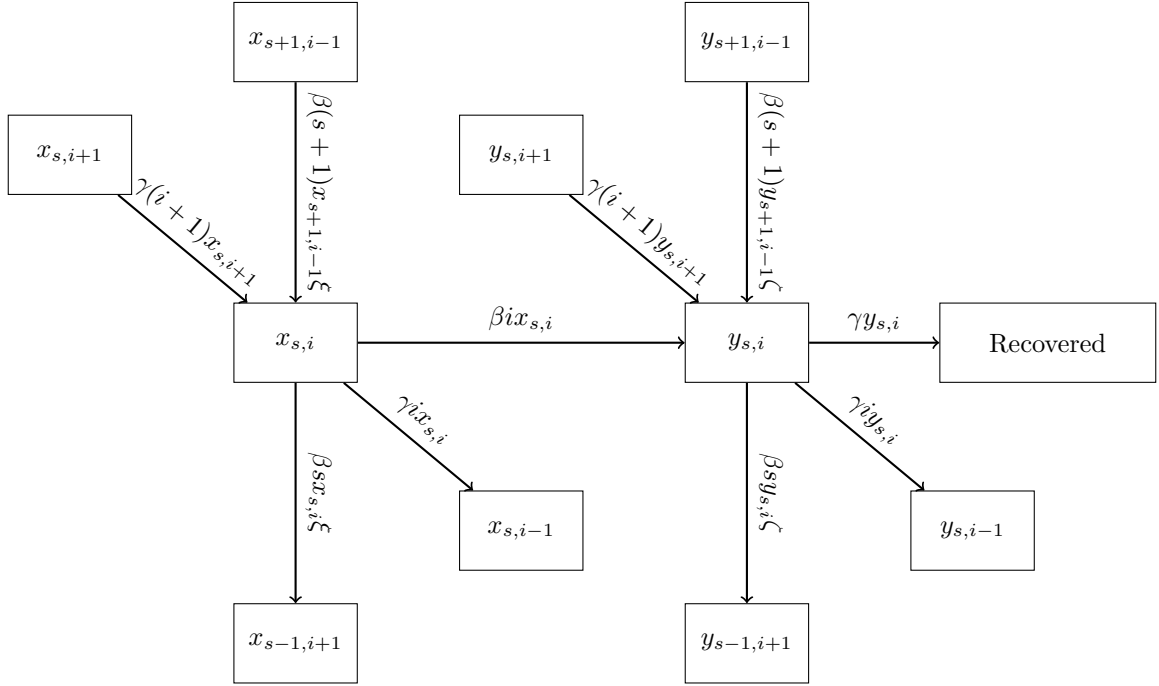
\begin{figure}
\begin{center}
\begin{tikzpicture}
\node [default] at (0,0) (xsi) {$x_{s,i}$};
\node [default] at (-3,2.5) (xsip1) {$x_{s,i+1}$};
\node [default] at (0,-4) (xsm1ip1) {$x_{s-1,i+1}$};
\node [default] at (0,4) (xsp1im1) {$x_{s+1,i-1}$};
\node [default] at (3,-2.5) (xsim1) {$x_{s,i-1}$};

\node [default] at (6,0) (ysi) {$y_{s,i}$};
\node [default] at (3,2.5) (ysip1) {$y_{s,i+1}$};
\node [default] at (6,-4) (ysm1ip1) {$y_{s-1,i+1}$};
\node [default] at (6,4) (ysp1im1) {$y_{s+1,i-1}$};
\node [default] at (9,-2.5) (ysim1) {$y_{s,i-1}$};

\node [wide] at (10,0) (recovered) {Recovered};

\path [->, thick, sloped, above] (xsip1) edge node {$\gamma (i+1) x_{s,i+1}$} (xsi);
\path [->, thick, sloped, above] (xsp1im1) edge node {$ \beta(s+1) x_{s+1,i-1}\xi$} (xsi);
\path [->, thick, sloped, above] (xsi) edge node {$\beta s x_{s,i}\xi$} (xsm1ip1);
\path [->, thick, sloped, above] (xsi) edge node {$\gamma i x_{s,i}$} (xsim1);

\path [->, thick, sloped, above] (ysip1) edge node {$\gamma (i+1) y_{s,i+1}$} (ysi);
\path [->, thick, sloped, above] (ysp1im1) edge node {$\beta (s+1) y_{s+1,i-1}\zeta$} (ysi);
\path [->, thick, sloped, above] (ysi) edge node {$\beta s y_{s,i}\zeta$} (ysm1ip1);
\path [->, thick, sloped, above] (ysi) edge node {$\gamma i y_{s,i}$} (ysim1);

\path [->, thick, sloped, above] (xsi) edge node {$\beta i x_{s,i}$} (ysi);
\path [->, thick, sloped, above] (ysi) edge node {$\gamma y_{s,i}$} (recovered);
\end{tikzpicture}
\end{center}
\caption{\textbf{The flow diagram underlying the effective degree model
  of~\cite{lindquist}}.  We include just the fluxes involving the
  $x_{s,i}$ or $y_{s,i}$ compartments.  Fluxes between other
  compartments exist but are not included (they can be deduced by changing $s$ and $i$ and using this figure).  Given a susceptible
  individual $u$, if $v$ is a susceptible partner of $u$, the variable
  $\xi$ is the expected number of infected partners of $v$.  Given an
  infected individual $u$, if $v$ is a susceptible partner of $u$,
  then $\zeta$ is the expected number of infected partners of $v$
  (including $u$).  We have made some small changes to the variables
  as used by~\cite{lindquist}: Their $G = \beta \xi$ and their $H =
  \beta \zeta$.}
\label{fig:lindquist}
\end{figure}

Following figure~\ref{fig:lindquist}, the final equations are 
\begin{align}
\dot{x}_{s,i} &= -\beta i x_{s,i} + \gamma \left[(i+1) x_{s,i+1} - i
  x_{s,i}\right] + \beta \xi \left[(s+1)x_{s+1,i-1}-sx_{s,i} \right]\label{eqn:lindquistalpha}\\
\dot{y}_{s,i} &= \beta i x_{s,i} - \gamma y_{s,i} + \gamma \left[
  (i+1) y_{s,i+1} - i y_{s,i} \right] +\beta  \zeta \left[
  (s+1)y_{s+1,i-1}-sy_{s,i} \right]\\
\xi &= \frac{\sum_{k_e=1}^M \sum_{s+i=k_e} s i x_{s,i}}{\sum_{k_e=1}^M
 \sum_{s+i=k_e}s x_{s,i}}\\
\zeta &= \frac{\sum_{k_e=1}^M \sum_{s+i=k_e} i^2 x_{s,i}}{\sum_{k_e=1}^M
 \sum_{s+i=k_e}i x_{s,i}}\\
S &= \sum_{k_e=1}^M \sum_{s+i=k_e} x_{s,i}\\
I &= \sum_{k_e=1}^M \sum_{s+i=k_e} y_{s,i} \\
R &= N-S-I \, .\label{eqn:lindquistomega}
\end{align}
This is a system with $2(M+1)^2$ differential equations where $M$ is the maximum degree.  Careful consideration can remove about half of these equations since the $\dot{x}_{s,i}$ equations do not depend on the $y_{s,i}$ terms.  So we can sum the $\dot{y}_{s,i}$ equations to arrive at a single equation for $\dot{I} = \beta \sum_{k_e=1}^M \sum_{s+i=k_e}  i x_{s,i} - \gamma I$.  So we can still calculate $S$, $I$, and $R$ if we reduce this system to $(M+1)^2 + 1$ differential equations.  The basic effective degree model was developed by~\cite{lindquist} in large part to study SIS disease, and this simplification would not be appropriate in the SIS context.

As before, we may switch to intensive quantities by dividing the $x$ and $y$ variables by $N$.

\subsubsection{The compact effective degree model}

A related model~\cite{ball:network_eqns}, uses a different classification to identify ``effective partnerships''.  As above, the model ignores partnerships involving recovered individuals.  However, we also assume that we can observe when a transmission occurs in a partnership.  So we can observe if a partnership was the original source of infection for either individual and we can observe those partnerships in which one infected individual ``transmits'' to its partner but fails to cause infection because the partner is already infected.  If a partnership has not transmitted infection and neither individual is recovered, then it is an effective partnership.

\begin{figure}
\begin{center}
\begin{tikzpicture}
\node [narrow] at (-1.8,0) (xip1) {$x_{j+1}$};
\node [narrow] at (5.4,0) (xim1) {$x_{j-1}$};
\node [narrow] at (1.8,0) (xi) {$x_j$};
\node [narrow] at (0,-3) (yi) {$y_j$};
\node [narrow] at (3.6,-3) (yim1) {$y_{j-1}$};
\node [narrow] at (-3.6,-3) (yip1) {$y_{j+1}$};
\node [wide] at (0,-6) (rec) {Recovered};

\path [->, thick,sloped,above] (xip1.315) edge node {$\beta \I (j+1)x_{j+1}$} (yi.90);
\path [->, thick, bend left, above] (yip1.45) edge node {$\beta (j+1)y_{j+1}$}  (yi.135);
\path [->, thick,above] (yip1.0) edge node {$\beta \I (j+1)y_{j+1}$}  (yi.180);
\path [->, thick, bend right,below] (yip1.315) edge node {$\gamma \I (j+1) y_{j+1}$}  (yi.225);
\path [->, thick, bend left, above] (yi.45) edge node {$\beta jy_j$}  (yim1.135);
\path [->, thick,above] (yi.0) edge node {$\beta \I j y_j$}  (yim1.180);
\path [->, thick, bend right, below] (yi.315) edge node {$\gamma \I j y_j$}  (yim1.225);
\path [->, thick, sloped, above] (yi) edge node  {$\gamma y_j$} (rec);

\path [->, thick,above] (xip1) edge node {$\gamma \I (j+1)x_{j+1}$} (xi);
\path [->, thick, sloped,above] (xi.315) edge node {$\beta \I j x_j$} (yim1.90);
\path [->, thick,above] (xi) edge node {$\gamma \I jx_j$} (xim1);


\end{tikzpicture}
\end{center}
\caption{\textbf{The flow diagram underlying the model
  of~\cite{ball:network_eqns}}.  Only the fluxes into and out of $y_j$
  and $x_j$ are included.  Fluxes between other compartments exist,
  but are not included.  An effective partnership is eliminated if the partner
  recovers or if it transmits infection in either direction.  The
  quantity $\I$ represents the probability an effective partnership joins an
  individual with an infected partner.}
\label{fig:ball}
\end{figure}
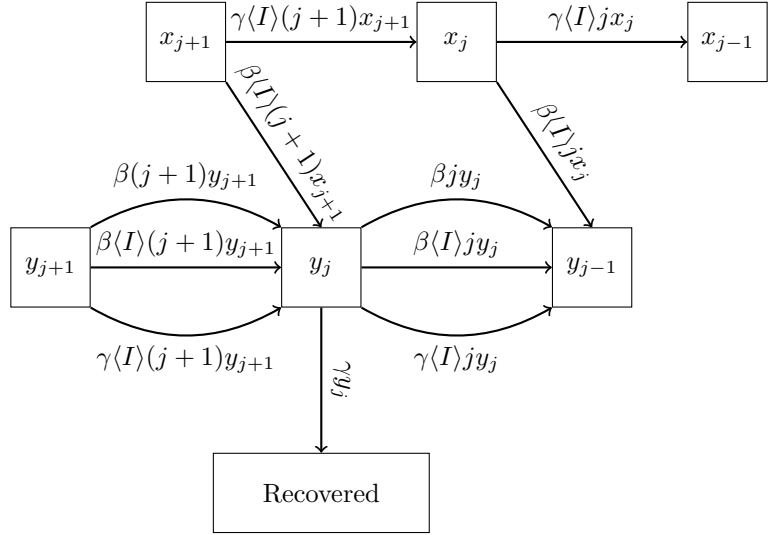

We now add the pairs closure assumption: Given a susceptible individual $u$ the probability that an effective partner $v$ is susceptible or infected is independent of any information we have about $u$ or the status of any other partners of $u$.  As a result, the probability that a susceptible individual $u$ has $s$ effective susceptible partners and $i$ effective infected partners can be deduced by knowing the probability of having $s+i$ effective partners and the independent probability that a partner is infected.  Thus rather than having $x_{s,i}$ we stratify into groups $x_j$ where $j$ is the number of effective partners of a susceptible individual.  We refer to this model as ``compact'' because we have just a single subscript to track.  We note that the ``reduced'' model introduced below is of similar complexity to this one.  

To derive the equations of~\cite{ball:network_eqns} we additionally assume that the probability an effective partner of an infected individual $u$ is infected is the same as it would be if $u$ were susceptible.  That is, we make the strong pair closure assumption.

We can calculate $\I$, the probability that an effective partner is infected.  A little consideration will show that this is $\I = \sum_j j y_j/ \sum_j j(y_j+x_j)$.
Thus on average for susceptible individuals with $j$ effective partnerships, we
conclude that the rate of becoming infected is $\beta j\I$.  Similarly
the rate at which such an individual loses an effective partner because the partner
recovers is $\gamma j \I$.  The flow diagram in
figure~\ref{fig:ball} shows the relevant transitions.  The resulting
equations are
\begin{align}
\dot{x}_j &= \gamma \I \big ( (j+1)x_{j+1}-jx_j \big) - \beta \I
jx_j \label{eqn:ball_alpha}\\
\dot{y}_j &= (\beta + \beta \I + \gamma \I) \big ( (j+1)y_{j+1}-jy_j
\big) + \beta \I(j+1)x_{j+1} - \gamma y_j\\
\I &= \frac{\sum_j jy_j}{ \sum_j j(x_j+y_j)} \\
S&= \sum_j x_j\\
I &= \sum_j y_j\\
R&= N-S-I \, . \label{eqn:ball_omega}
\end{align}
This system has significantly fewer equations than the basic effective degree model, just $2(M+1)$ differential equations where $M$ is the maximum degree.  These are the equations of~\cite{ball:network_eqns}.  We can eliminate $y_j$ from this system by noting that $\I$ depends on $y_j$ only through the sum $\sum_j j y_j$.  If we define $\lambda= \sum_j j y_j$ we have $\I = \lambda/(\lambda + \sum_j j x_j)$ and we find
\[
\dot{\lambda}  = \beta \I \left(\sum_j j(j+1) x_{j+1}\right) - [ \beta + \gamma + (\beta+\gamma)\I]  \lambda \, .
\]
So if we know the initial value of $\lambda$, we do not need to track the change in each individual $y_j$.  This substantially reduces the number of differential equations.  However, because we no longer have an individual equation for each $y_j$, we no longer have an equation for $I$.  We must add one differential equation back.  The simplest form is to take $S= \sum_j x_j$, \ $I = N-S-R$, and $\dot{R} = \gamma I$.  Thus we have $M+3$ differential equations.

The role of $\lambda$ is to count the number of effective partnerships that involve infected individuals (counting double if both are infected).  The role of $\sum_j j x_j$ is to count the number of effective partnerships that involve susceptible individuals (also counting double if both are susceptible).  So $\I = \lambda/ (\lambda + \sum_j j x_j)$ measures the probability that an effective partner is infected.  Thus the probability that a susceptible individual's effective partner is infected is $\I$.

\subsubsection{Reduced Effective Degree Model}
With a small modification of our definition of effective partnerships,  we can eliminate the need to track the number of partners infected individuals have altogether.  We add to our definition the statement that an effective partnership must have at least one susceptible individual.  With this definition, we will not need to track partnerships between infected individuals.  This new approach will not require the strong version of the pairs closure required for our derivation of the compact effective degree model.

We define $\nu$ to be the number of effective partnerships between susceptible and infected individuals.  Using this definition, $\I = \nu /\sum_j j x_j$.  There are several ways that $\nu$ can change.  Infection of a partner in a susceptible-susceptible pair increases $\nu$.  The contribution to $\dot{\nu}$ from this is $\sum_j\beta \I j x_j (j-1)(1-\I)$.  The $\beta \I j x_j$ gives the rate at which $x_j$ individuals are infected, and $(j-1)(1-\I)$ gives the expected number of new susceptible-infected partnerships resulting.  The value of $\nu$ decreases when the infected partner within a susceptible-infected partnership recovers or transmits to the susceptible partner.  This contributes $-(\beta+\gamma) \nu$ to $\dot{\nu}$.  Finally $\nu$ decreases when the susceptible member of a susceptible-infected partnership is infected from outside the partnership.  This contributes $- \beta\I^2 \sum_j j (j-1)x_j$ to $\dot{\nu}$.  We are left
\[
\dot{\nu} = -(\beta+\gamma) \nu + \beta \I (1-2\I)\sum_j j (j-1)  x_j  \, .
\]
Note that the derivation of $\nu$ does not rely on any independence assumption of partners of infected individuals.  This yields the new system
\begin{align}
\dot{x}_j &= \gamma \I \big ( (j+1)x_{j+1}-jx_j \big) - \beta \I
jx_j \label{eqn:ball_improved_alpha}\\
\dot{\nu} &= -(\beta+\gamma) \nu + \beta \I(1-2\I) \sum_j j (j-1)x_j\\
\I &= \frac{\nu}{\sum_j jx_j} \\
S &= \sum_j x_j\\
I &= N-S-R\\
\dot{R} &= \gamma I \, .\label{eqn:ball_improved_omega}
\end{align}
We again have $M+3$ equations, but we do not need to assume anything about the partners of infected individuals.  The derivation of this requires only the pairs closure.

Again, we may switch to intensive quantities by dividing all $x$ and variables and $\nu$ by $N$.

This system can be derived by adding the pairs closure to the basic effective degree model.  We demonstrate this in the appendix.

\subsection{Edge-based compartmental modeling}

We finally turn to a third approach which has been called the ``Edge-based compartmental modeling'' (EBCM) approach.  This approach relies on creating compartments of partnerships (or edges) rather than compartments of individuals.  It is also sometimes referred to as a ``pgf'' approach because the equations contain probability generating functions.  The ``pgf'' terminology may be more descriptive because the pairwise equations also are based on compartments of edges.  Nevertheless, we will stick to the ``EBCM'' terminology.  The method is described in~\cite{miller:ebcm_overview}.  Some of the definitions of~\cite{miller:ebcm_overview} may at first seem unusual, but this is done to simplify the resulting mathematics by eliminating the need to use conditional probability arguments.  We begin by discussing the variables used.  Then we derive the governing equations.

\subsubsection{The EBCM variables}

Under the assumption that the population-scale epidemic dynamics are deterministic, the probability a random individual is in a given state will match the proportion of the population in that state.  Thus our derivation will focus on calculating the probability a random individual is in each state.  

The pairs closure guarantees that as long as an individual $u$ is susceptible, we can treat its partners as independent.  Our goal is to calculate the probability that if $v$ is a partner of $u$, then at time $t$ no individual $x$ reachable from $u$ through $v$ has caused a chain of transmissions that reach $u$.  This is complicated somewhat because if $u$ is infected from another source, it can infect $v$, which short-circuits the potential chain of transmissions from $x$.  We can get around this by conditioning on the assumption $u$ has not been infected from another source, but the arguments are quickly full of conditional probability.  Instead we modify $u$ to prevent $u$ from transmitting to its partners if infected.  This prevents the transmission chains to $u$ from interacting with one another.  The status of partners of $u$ are independent.  This modification does not alter the probability that $u$ has a given status because it only changes things once $u$ is infected, and at that point the recovery time of $u$ is unaffected by any dynamics extrinsic to $u$.  This is discussed in more detail in~\ref{sec:notrans}.

When $u$ is a randomly chosen individual which is prevented from causing infections, we refer to $u$ as a \emph{test individual}.  We now are able to define our variables.  Our central variable is $\theta$, the probability that a partner $v$ of a test individual $u$ has not yet transmitted to $u$.  We break $\theta$ up into three variables $\theta = \phi_S+\phi_I+\phi_R$.  Each is the probability that $v$ has not yet transmitted to $u$ and is in the given state.

We can avoid the modification done in defining the test individual by using an alternate definition.  We choose $u$ uniformly at random from the population and $v$ as a uniformly chosen random partner of $u$.  Then $\theta$ is the probability that no transmission has occurred between the two given that either no transmission has happened or if transmission has happened, the first transmission was $v$ to $u$.  Although this alternative definition would lead to the same equations, for consistency with prior work, we will use the test individual assumption.  So $u$ will be prevented from transmitting.

\subsubsection{Equation derivation}

In deriving the equations, we use $S$, $I$, and $R$ to be the number of susceptible, infected, or recovered individuals to be consistent with the models discussed previously.  In contrast, in other work with the EBCM approach, these are typically taken to be the proportions.  

Let us now consider a randomly chosen individual $u$ in the population.  We define $S_k(0)$ to be the number of individuals which are initially susceptible and have degree $k$.  We define $\theta(t)$ to be the probability that a partnership to a test individual $u$ has not yet transmitted to $u$ by time $t$ assuming that $u$ was initially susceptible [so $\theta(0)=1$].  We assume that $\theta$ is independent of the degree of $u$.  Averaging over all $k$, the probability that $u$ is susceptible at time $t$ is thus $\sum_k \frac{S_k(0)}{N} \theta(t)^k$.  We define this function to be $\psi(\theta)$.  The number of infections is $S(t) = N\psi(\theta)$.  To find $R(t)$ and $I(t)$, the number that are recovered or infected respectively, at time $t$ we take $\dot{R} = \gamma I$, and $I = N-S-R$.  In the limit that $S_k(0)$ can be approximated as $NP(k)$ (that is, a negligible proportion of each degree initially infected), the function $\psi(\theta)$ is the probability generating function of the degree distribution.

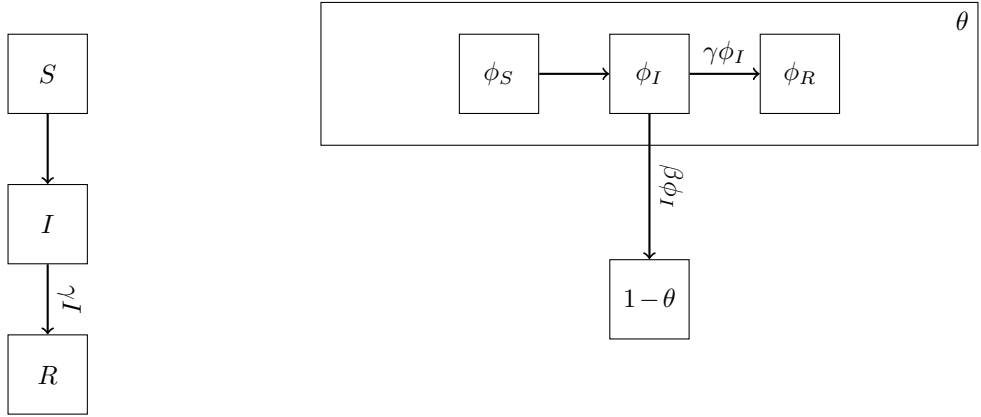
\begin{figure}
\begin{center}
\begin{tikzpicture}
\node [narrow] at (-2,0) (phis) {$\phi_S$};
\node [narrow] at (0,0) (phii) {$\phi_I$};
\node [narrow] at (2,0) (phir) {$\phi_R$};
\node [narrow] at (0,-3) (1mt) {$1-\theta$};
\node [bigS] at (0,0) (Btheta) {\parbox{8.5cm}{\hfill $\theta$\\[1cm]~}};

\path [->, thick,sloped,above] (phis) edge node {} (phii);
\path [->, thick,sloped,above] (phii) edge node {$\gamma \phi_I$} (phir);
\path [->, thick,sloped,above] (phii) edge node {$\beta \phi_I$} (1mt);

\node [narrow] at (-8,0) (S) {$S$};
\node [narrow] at (-8,-2) (I) {$I$};
\node [narrow] at (-8,-4) (R) {$R$};
\path [->, thick,sloped,above] (S) edge node {} (I);
\path [->, thick,sloped,above] (I) edge node {$\gamma I$} (R);

\end{tikzpicture}
\end{center}

\caption{\textbf{The flow diagram for the basic Edge-Based Compartmental
  Model}.  We do not calculate the flux from $\phi_S$ to $\phi_I$ or
  from $S$ to $I$ directly because we calculate $\phi_S$ and $S$
  directly.  The relevant fluxes can be found by differentiating
  $\phi_S$ or $S$.  Using the relative fluxes from $\phi_I$, we can
  calculate $\phi_R$ in terms of $\theta$.}
\end{figure}

To complete our system, we need to know $\theta$.  This is the probability that a partner $v$ of $u$ has not yet transmitted to $u$ given that $u$ was initially susceptible.  It is straightforward to see that $\dot{\theta} =-\beta \phi_I$ because $\theta$ is decreased by transmissions, transmissions happen along partnerships at rate $\beta$, and $\phi_I$ is the probability transmission has not yet happened and $v$ is infected.  Our goal will be to find $\phi_I$ in terms of $\theta$ to reduce our system to just this single differential equation.

We begin by calculating $\phi_S$ in terms of $\theta$.  If $v$ was initially susceptible, the probability it has degree $k$ is $k S_k(0)/ \sum_k k S_k(0)$.  Assuming it was initially susceptible with degree $k$, the probability that it is still susceptible is $\theta^{k-1}$ (each partnership other than the one with $u$ has a chance to transmit to $v$)\footnote{To help make the derivation clearer, we can make a test-individual like assumption so that $v$ can only transmit to $u$ and no other partners.}.  So the probability that $v$ is susceptible at time $t$ is $\phi_S(0) \sum_k k S_k(0) \theta^{k-1}/ \sum_k k S_k(0) = \phi_S(0) \psi'(\theta)/\psi'(1)$.

We now find $\phi_R$.  The rate at which an infected neighbor recovers
is $\gamma$.  The rate at which an infected neighbor transmits to $u$
is $\beta$.  Thus, we find that $\dot{\phi}_R =
-\gamma\dot{\theta}/\beta$.  Taking $\phi_R(0)$ as given and
$\theta(0)=1$, we conclude $\phi_R(t) = \phi_R(0) + \gamma
(1-\theta)/\beta$.

We finally conclude that $\phi_I = \theta - \phi_S(0)
\psi'(\theta)/\psi'(1) - \gamma(1-\theta)/\beta - \phi_R(0)$.  Thus we
arrive at
\begin{align}
\dot{\theta} &= - \beta \theta + \beta \phi_S(0) 
\frac{\psi'(\theta)}{\psi'(1)} + \gamma(1-\theta) + \beta \phi_R(0) \label{eqn:EBCMalpha}\\
S&= N\psi(\theta) \\
I &= N-S-R \\
\dot{R} &= \gamma I  \, . \label{eqn:EBCMomega}
\end{align}
The single differential equation for $\theta$ drives all the dynamics, but if we want $S$, $I$, and $R$ as well, we need to include additional equations.  We require just two differential equations.

Changing to intensive rather than extensive variables has no impact on the equation for $\theta$ and simply divides $S$, $I$, and $R$ by $N$.

\section{Model Selection}

\begin{figure}
\begin{tikzpicture}
\node [rounded] at (0,1.) (global) {Global exact (unclosed) model};
\node [rounded] at (-6,0) (BPW) {Basic Pairwise Model};
\node [rounded] at (6,0) (BED) {Basic Effective Degree Model};
\node [rounded] at (-6,-2.5) (EBCM) {Edge-Based Compartmental Model};
\node [rounded] at (0,-2.5) (CPW) {Compact Pairwise Model};
\node [rounded] at (6,-2.5) (RED) {Reduced Effective Degree Model};
\node [rounded] at (6,-5) (CED) {Compact Effective Degree Model};

\path [->, sloped, above, >=triangle 90] (global) edge node {Triples Closure} (BPW.0);
\path [->, sloped, above, >=triangle 90] (global) edge node {Star Closure} (BED.180);
\path [->, sloped, above, >=triangle 90] (BPW) edge node {Pair Closure} (CPW.135);
\path [->, sloped, above, >=triangle 90] (BED) edge node {Pair Closure} (CPW.45);
\path [<->, >=triangle 90] (EBCM) edge node {} (CPW);
\path [<->, bend right, >=triangle 90] (EBCM.0) edge node {} (RED.180);
\path [<->, >=triangle 90] (CPW) edge node {} (RED);
\path [->, right, >=triangle 90] (RED) edge node {Strong Pair Closure} (CED);
\end{tikzpicture}
\caption{\textbf{The hierarchy of models}.  Labeled edges denote the additional
  assumptions required to find the mathematically simpler model.
  Unlabeled (bidirectional) edges imply that the two models are
  equivalent and each can be derived from the other.  The unclosed
  exact model will reduce to either the basic effective degree model
  of~\cite{lindquist} or the basic pairwise
  model~\cite{eames:pair} depending on the closure used.  Each
  of these then reduces to the compact pairwise model (as modified from~\cite{ball:network_eqns}).  The compact pairwise model, the  reduced 
  effective degree Model, and the Edge-based Compartmental model~\cite{miller:ebcm_overview} are all equivalent.  Using the strong version of the pair closure, we can derive the compact effective degree model which also tracks information about infected individuals and their partners.}
\end{figure}
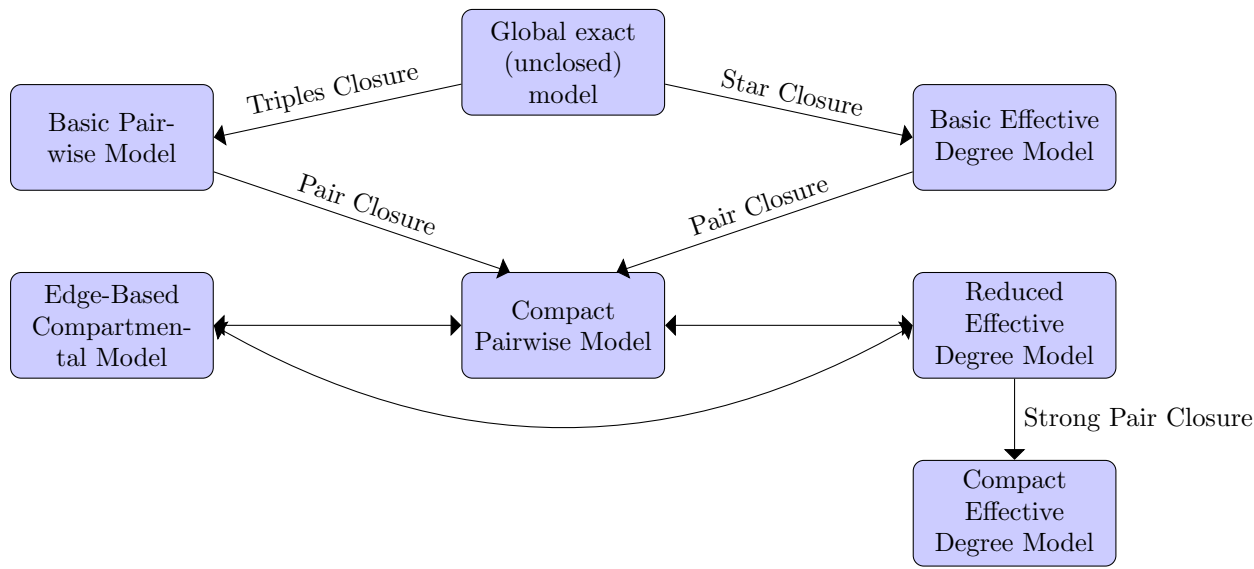

\subsection{Relationship of models}

The models are based on different assumptions about the independence
of partners.  Although the assumptions are formally different, the
distinction is in a sense unnatural.  If the initial infections are
randomly selected, all the assumptions are satisfied, and the models
give identical predictions.  If the disease is introduced as a few
infections, and allowed to spread until enough infections have
occurred that the dynamics are deterministic, then all the assumptions
will be satisfied, and the models give identical predictions
(technically it is possible that stochastic effects cause one
assumption to be valid but not another, but this is typically insignificant).

In order for the assumptions of any of the models to be violated, this must be built into the initial condition.  Somehow the initial infections must be chosen based on who their neighbors are.  Even if the initial condition fails to satisfy assumptions of some of the models, as time progresses information about the initial condition is lost.  As the disease spreads, the violated assumptions may become valid (at least approximately) and the corresponding model is accurate.

Formally speaking, some of the models make stronger assumptions than
others, and we prove in the appendix that it is possible to derive
these models from the models having weaker assumptions by strengthening the asumptions.  In the
case in which we if we take a random susceptible individual  $u$ we
find that the statuses of partners of $u$   are independent of the
degree of $u$ and the status of one another, then we find that all of
the models are appropriate.  We can reduce all of the models to the
EBCM model, having just a single ODE driving the dynamics (with a second ODE needed for bookkeeping).

\begin{table}
\begin{center}
\begin{tabular}{|c|c|c|c|}
\hline
\parbox{0.2\textwidth}{\T{}Model\B{}}
& \parbox{0.22\textwidth}{\T{}Minimal number of Differential
  Equations\B{}} & \parbox{0.22\textwidth}{\T{}Analytic final size calc?\B{}} & \parbox{0.2\textwidth}{\T{}Early
growth calc?}\\ \hline
\parbox{0.2\textwidth}{\T{}Basic pairwise model\B{}} & $2K^2+K+1$ & No & \parbox{0.2\textwidth}{\T{}$\order(K^2 \times K^2)$
eigenvalue problem\B{}}\\ \hline
\parbox{0.2\textwidth}{\T{}Compact pairwise model\B{}} & $K+3$ &
No & Analytic calculation\\ \hline
\parbox{0.2\textwidth}{\T{}Basic effective degree model\B{}} &
$(M+1)^2+1$ & No  & \parbox{0.2\textwidth}{\T{}$\order(M^2\times M^2)$
eigenvalue problem\B{}}\\ \hline
\parbox{0.2\textwidth}{\T{}Compact  effective degree model\B{}} &
$M+3$ & No & \parbox{0.2\textwidth}{\T{}$\order(M\times M)$ eigenvalue problem.\B{}}\\ \hline
\parbox{0.2\textwidth}{\T{}Reduced effective degree model\B{}} & $M+3$ & No &Analytic calculation.\\ \hline
\parbox{0.2\textwidth}{\T{}Edge Based Compartmental model\B{}} & $2$ &
Yes & Analytic calculation.
\\\hline
\end{tabular}
\end{center}
\caption{\textbf{Comparison of the models}.  The number of equations needed and computational/analytic tractibility for a population having $K$ distinct
  degrees with maximum degree $M\geq K$.}
\end{table}

We have discussed six distinct models of epidemic spread in static
configuration model networks.  These models have varying mathematical
and conceptual complexity.  A natural question is which model is
appropriate for a given application.

\subsection{Model robustness to assumptions}

All of these models make some form of simplifying assumption (a
closure) in the derivation.  At the heart of each of the assumptions
is some sort of independence assumption.  Each assumption can be shown
to be valid if there is no information content in certain scales.
Since the epidemic is a stochastic process, whatever information may
be imposed by the initial conditions is lost over time.  We anticipate
that all models should behave similarly at later time, and should fit
observed dynamics closely.

\subsection{Robustness to Configuration Model assumptions}
These models have been derived by assuming that the population is a
static Configuration Model network.  This assumption is unrealistic:
individuals change partners, and there may be correlations between the
degrees of partners.  

\subsubsection{Degree correlations}
If we allow for correlations to exist between the degrees of
individuals, then the effective degree approaches will have to keep
extra information.  It is not enough to know how many effective partnerships an
individual has, we need to know how many partners the individual began
with in order to know the risk from those effective partnerships.  This
significantly increases the dimensionality of these models.

The basic pairwise model approach remains valid if degree correlations exist.  The triples closure can hold even if the distribution of partners' degrees depends on an individual's degree.  If it holds, then the approach used to derive the basic pairwise model will yield a usable system of equations~\cite{eames:pair}.

The simplifications that go into deriving the compact pairwise model
from the basic pairwise model fail if an individual's degree affects
its choice of partner.  So we will not be able to use this model
(though similar simplifications may be possible, with $\I$ becoming a function of degree).

The Edge-Based Compartmental Model presented here was derived assuming
that $\theta$, the probability a partner of individual $u$ has not transmitted to
it, is independent of the degree of $u$.  Clearly if there
are degree correlations, this will not be true.  However, depending on
how those correlations occur, a modification of the EBCM approach may
work.  Details of how this can be done are
in~\cite{miller:ebcm_structure}.  It requires allowing $\theta$ to be
a function of $k$ as well as $t$.  So rather than a single equation
for $\theta$, we find $K$ equations where $K$ is the number of
distinct degrees.

\subsubsection{Dynamic networks}
There are multiple ways we can include partner turnover in the model.
One of the most obvious is to allow an individual's partnerships to
end and be replaced by new partnerships.  Another is to allow
individuals to have intrinsic rates of creating new partnerships
while they end existing partnerships independently.  Other models are
clearly possible.

Again, the effective degree models would require a significant
increase in dimension to account for partnership turnover.  We would
need to track the number of ineffective partnerships in some way.

The Edge-Based Compartmental Modeling approach is well-suited for
many dynamic networks.  Examples were developed 
in~\cite{miller:ebcm_overview}, and in~\cite{miller:ebcm_hierarchy} it
was shown that a hierarchy of models could be constructed based on the
assumptions about partner turnover.  It was shown that a widely-used model~\cite{may:hivdynamics, may:dynamics,moreno,pastor-satorras:scale-free}
that assumes that at every moment partnerships are re-selected
randomly arises as a special case of this hierarchy, and indeed the
standard mass action model can be recovered in appropriate limits.
One particular case that should be noted is that it is possible to
model some forms of ``serosorting'' where individuals actively select
partners of similar infection status.

Depending on how the partnership turnover is modeled, the compact pairwise model may be robust enough to accommodate it.  The EBCM approach for dynamic networks requires introducing additional variables that are equivalent to $[SS]$, $[SI]$, $[SR]$ and related quantities.  So we hypothesize that an attempt to adapt the compact pairwise model to a dynamic network is likely to lead to the same equations found by the EBCM approach.

It is relatively simple to adapt the basic pairwise model to most
dynamic networks considered in the generalizations of the EBCM
model of~\cite{miller:ebcm_overview}.

\subsection{Final size relations and early growth}
\subsubsection{Early growth}
One of the most important questions to answer about a disease is
whether it is capable of causing an epidemic in a given population.
To answer this, we typically look at $\Ro$, the average number of
infections caused by an early infectious individual, and try to find
conditions under which $\Ro<1$.  However, for models which calculate
the dynamic growth of an epidemic, it is difficult to extract $\Ro$
from the model.  Instead, we look at an equivalent measure, the early
growth rate.  

We assume the epidemic is initialized with an infinitesimally small
proportion infected and grows proportionally to $e^{rt}$, and try to
calculate the value of $r$.  If $r>0$, epidemics are possible.
Otherwise they are not.

For the basic pairwise model, we take the $[S_kI_{k'}]$ equations from
equations~\eqref{eqn:basicalpha}--\eqref{eqn:basicomega}.
Because we assume the amount of infection is small, the
$[IS_k][S_kI_{k'}]$ term is negligibly small.  The other terms however
are not negligible.  The resulting system yields a $K^2 \times K^2$
eigenvalue problem whose most positive (or least negative) eigenvalue determines the growth rate.

For the compact pairwise model,
equations~\eqref{eqn:compactalpha}--\eqref{eqn:compactomega}, we can
note that so long as the amount of infection is very small, the
denominator of $\I$ may be treated as constant, so $\I$ is proportional
to $[SI]$.  Substituting this into the equation for $[SI]$, we discard
the $\beta \kex \I [SI]$ term.  Assuming that all but an infinitesimal
proportion of the population is susceptible (so $[SS]\I = [SI]$), we
arrive at $\dot{[SI]} = \beta \kex(0)  [SI] -
(\beta+\gamma) [SI]$, and so we conclude that 
\[
r = \beta \kex(0) - (\beta + \gamma) \, .
\]

For our first effective degree model, equations~\eqref{eqn:lindquistalpha}--\eqref{eqn:lindquistomega}, the original
paper~\cite{lindquist} showed that the early growth rate could be
derived from a $(M+1)^2-1 \times (M+1)^2-1$ eigenvalue problem.

For the compact effective degree model,
equations~\eqref{eqn:ball_alpha}--\eqref{eqn:ball_omega}, we can convert
the $y_j$ equations into an $M\times M$ eigenvalue problem (neglecting $y_0$).  If we use the reduced effective degree model equations~\eqref{eqn:ball_improved_alpha}--\eqref{eqn:ball_improved_omega}, we get 
\begin{align*}
\dot{\nu} &= -(\beta+\gamma)\nu + \beta\I(1-2\I) \sum_jj(j-1) x_j \\
&= -(\beta+\gamma)\nu + \beta\frac{\nu}{\sum_jjx_j}(1-2\I) \sum_jj(j-1) x_j \\
&= -(\beta+\gamma) \nu + \beta (1-2\I)\nu\frac{\sum_jj(j-1)x_j}{\sum_jjx_j} \, .
\end{align*}
Assuming $\nu$ and $\I$ are both small, we have at leading order
\[
\dot{\nu} = \left(-(\beta+\gamma) + \beta \frac{\sum_jj(j-1)x_j}{\sum_jjx_j} \right)\nu \, .
\]
At $t=0$, the value of $\sum_jj(j-1)x_j/\sum_jj x_j$ corresponds to $\kex(0)$, so the resulting growth rate corresponds to the previous value.

Finally, for the EBCM model, we set $\theta = 1-\epsilon e^{rt}$ and
solve for $r$.  The equation for $\dot{\theta}$ becomes
\[
-r \epsilon e^{rt}  = -\beta + \beta \epsilon e^{rt} + \beta
\frac{\psi'(1) - \epsilon e^{rt}\psi''(1)}{\psi'(1)} + \gamma \epsilon
e^{rt} \, .
\]
After some rearranging, 
\[
r =  \beta \frac{\psi''(1)}{\psi'(1)} - (\beta+\gamma) \, .
\]
We note that $\kex(0) = \psi''(1)/\psi'(1)$, so the analytic formulae
we derive with the different models are identical.

\subsubsection{Final size}

To search for a final size relation, we set all derivatives to $0$ and
look to see if we can find an equation for the final size.  The only
case in which a clear solution exists is the EBCM model.  We can
trivially arrive at an equation which gives the fixed point of
$\theta$.  Using this fixed point, we immediately have
$S=\psi(\theta)$, and taking $I=0$, we conclude that $R = 1-\psi(\theta)$.

\section{Current Challenges}

There are a number of questions that remain unanswered for epidemic
spread in networks.  Two that appear both important and
difficult are the spread of disease through non-Configuration Model
networks (in particular clustered networks) and the
spread of SIS disease, that is diseases in which individuals return to
a susceptible state.  In both cases the challenge is similar: The
independence assumptions made in our earlier models are not valid.

\subsection{Clustered networks}
Social contact networks violate the assumptions of a Configuration Model
networks in a number of ways.  One of the clearest violations which is
particularly important for the modeling of infectious disease spread
is that social contact networks often have clustering: The partners of an individual
are likely to be partners of one another.  

As a consequence, all of the closure approximations applied above fail.  There have been a number of attempts to study disease spread in clustered networks, but most rely on some sort of approximation~\cite{serrano:prl,britton:cluster,eames:clustered,gleeson:clustering_effect,mcbryde:cluster_threshold,melnik:unreasonable,miller:RSIcluster}. A select few avoid such approximation, but only by defining their way out of it: If we restrict our attention to clustered networks which contain very specific motifs, then we can make analytic progress~\cite{karrer:random_clustered,miller:random_clustered,newman:cluster_alg,volz:clustered_result}.  However, these networks have very specific restrictions.  Despite the specific structure, the number of equations in the corresponding ODE systems increases and it becomes difficult to extract analytic results from this.

We foresee that closures at higher structure levels are
possible~\cite{house:motif} and that the appropriate closures must
preserve the structure of cycles within the population.  However, this
requires knowledge of the motif structure, and may require specialized
generation algorithms.

It is possible to generate clustered networks with identical
cluster ``density'' but very different large scale structure, and
hence very different epidemic dynamics~\cite{green:clustered}.  Thus
clustering alone is not a sufficient metric to describe the network
structure.  Different motif types (\emph{e.g.}, full connected
triangles, squares, and fully connected squares with a missing
diagonal) may lead to similar levels of clustering but different
overall epidemic dynamics.

Our goal of course is a model which can take an arbitrary social
network and accurately predict the dynamics of an epidemic spreading
through the network.  However, in the immediate future, we foresee that
any approximate closure models will generally be developed on a
case-by-case basis, lacking the generality of SIR models on
Configuration Model networks.

\subsection{SIS models}

The key simplification underlying the EBCM model was that in calculating the status of individual $u$, we can ignore the possibility of transmissions from $u$ to its partners.  For an SIS epidemic, this assumption fails.  In a Configuration Model SIR epidemic, the partners of an individual $u$ are independent until $u$ becomes infected.  Then the partners are no longer independent; however, they can no longer influence the status of $u$ because it is infected.  So although dependence of partners exists, it has no effect on $u$.  In an SIS epidemic, once infected, $u$ can return to a susceptible state and be reinfected.  The dependence of partners of $u$ is no longer negligible.

Indeed, the fact that $u$ can alter its future infection probability appears to underlie some initially surprising results~\cite{chatterjee:SIS}.  It has been proven that certain networks are able to sustain an epidemic even though simple mean-field models suggested they could not.  The apparent reason is that high degree individuals maintain islands of infection about them by being continually reinfected (and occasionally managing to have infection spread along paths to other high degree individuals).

Our closure assumptions fail once an individual has an opportunity to infect its neighbors, so we must adapt our approaches. One of our challenges in modeling SIS disease is to identify appropriate closures which satisfy the dynamics.  It should be noted that although the first effective degree model, equations~\eqref{eqn:lindquistalpha}--\eqref{eqn:lindquistomega}, does not compare particularly well to the other models in number of equations, this is an unfair comparison.  It is based on a model developed for application to SIS epidemics.  For SIS epidemics it is likely to outperform generalizations of the more specialized SIR models because the star closure is expected to be more accurate than the other closures in the SIS case.

\section{Discussion}
There are multiple approaches used to model the dynamic spread of SIR
epidemics in networks.  These approaches have been developed under the
assumption of a Configuration Model network.  They perform well and give
identical results under reasonable initial conditions.  We have
identified the precise conditions under which the models are
equivalent.  The models have varying complexity, with the simplest
analytic model being the Edge-based Compartmental Model of~\cite{miller:volz,miller:ebcm_overview}

Some of the approaches can be adapted to population which are
assortatively or disassortatively mixed.  Other features such as
dynamic partnerships or link weights can also be accounted for by some
of these approaches.  

We have discussed two important restrictions to the approaches developed so
far.  These models all make some degree of assumption of independence
of neighbors of a susceptible individual $u$.  In a network with short
cycles, this assumption breaks down.  Similarly for an SIS disease,
infection could travel from one partner of $u$ to another through
$u$.  Even once $u$ returns to susceptible, we can no longer treat
the partners as truly independent.  We can find very special
classes of clustered networks for which our SIR models can be adapted
(by assuming very special motifs exist), but the number of equations
grows rapidly, and the approach does not appear to be generally
applicable.  For SIS epidemics, there does not appear to be any
analytic exact model, except in the limit of partnerships having
negligible duration.

\section*{Acknowledgments}
JCM was supported in part by 1) the RAPIDD program of the Science and Technology Directorate, Department of Homeland Security and the Fogarty International Center, National Institutes of Health, and 2) the Center for Communicable Disease Dynamics, Department of Epidemiology, Harvard School of Public Health under Award Number U54GM088558 from the National Institute Of General Medical Sciences.  The content is solely the responsibility of the authors and does not necessarily represent the official views of the National Institute of General Medical Sciences or the National Institutes of Health. The funders had no role in study design, data collection and analysis, decision to publish, or preparation of the manuscript.

\appendix
\section{Further EBCM details}
The EBCM model begins with the pairs closure.  We make an observation that if the dynamics are deterministic, then given the initial conditions, the probability a random individual is susceptible, infected, or recovered is equal to the proportion of the population in the corresponding state.  This allows us to switch from the question of what proportion of the population is in a given state to an equivalent question of what is the probability an individual is in a given state.  This new, equivalent question is more amenable to probabilistic arguments.

By noting that the probability an individual is in a given state is not affected by whether it transmits to its partners once infected, we can shift to yet another equivalent question which allows us to assume the partners of the random individual are independent.  This is the key conceptual step in the derivation.

\subsection{Variable choice}
In this section we motivate the variable choice used in the EBCM approach.  We begin by following a few steps from an early version of the EBCM approach~\cite{volz:cts_time}.  When we consider a susceptible individual $u$ with $k$ partners, the probability $p_I$ that one partner is infected (conditional on $u$ being susceptible) is independent of any other partners and is independent of $k$.  If $\beta$ is the transmission rate and $S_k(t)$ is the number of susceptibles of degree $k$, we conclude that $\dot{S}_k = -\beta k p_I$.  From this, we can conclude that $S_k = S_k(0) e^{k(-\beta \int_0^t p_I(\tau) \, \mathrm{d}\tau)}$.  Setting $\theta(t) = e^{-\beta\int_0^t p_I(\tau)\, \mathrm{d}\tau}$, we conclude that $S_k(t) = S_k(0) \theta(t)^k$.   Then $S(t) = \sum_k S_k(0)\theta(t)^k$.  The variable $\theta$ can be roughly interpreted as the probability that a partner $v$ of $u$ has not transmitted to $u$.   Taking the limit where $S_k(0) \approx NP(k)$, we arrive at $S(t) = N\psi(\theta(t))$ where $\psi(\theta) = \sum_k P(k) \theta^k$ is the probability generating function of the degree distribution. 

The fact that $S_k(t)$ is proportional to $\theta(t)^k$ suggests that $\theta$ is somehow a measure of the independent probability that one partner of $u$ has not yet transmitted to $u$.  However, it is not immediately obvious how to precisely define the relevant quantity that is $\theta$.  If we think of ``transmission'' as being the first time that $u$ receives an infectious dose, then the probabilities are dependent.  Once one individual has transmitted, the others cannot --- indeed the definition of $p_I$ used above becomes meaningless as soon as a transmission has occurred.  There are multiple ways to get around this.  In~\cite{volz:cts_time}, the approach was to take $\theta$ to be the probability an initially susceptible degree $1$ individual is still susceptible, and then show that $S_k(t)$ is proportional to $\theta(t)^k$.  Then it is possible to derive equations for $p_I$ with (considerable) effort.  An alternate approach introduced by~\cite{miller:volz} and extended in~\cite{miller:ebcm_overview} is the one we will use here.

We first define ``transmission'' to be as we did in the compact effective degree model.  An infected individual can still receive a transmission from another infected individual.  However, it has no effect in terms of infection.  This definition is not enough to create the independence of different partners that we need to proceed.  The possibility exists that $w$ will transmit to $u$ which will transmit to $v$ which will transmit back to $u$.  So transmission from $v$ to $u$ is more likely if there has been transmission from $w$ to $u$.  What we actually want is to measure the probability that there is at least one individual which is reachable from $u$ through $v$ that begins infected and sparks a chain of transmissions through $v$ that would reach $u$ by time $t$.  Having a $w$ to $u$ to $v$ back to $u$ chain would short circuit this.  So we need to somehow condition on not having this short circuit.  There is a lot of conditional probability in the resulting definition, so it is cumbersome to work with.  We can eliminate the need for conditional probability arguments by modifying $u$ at the outset.  The individual $u$ can receive transmissions, but cannot transmit to its partners.  This eliminates the $w$ to $u$ to $v$ chain without affecting the ability of infection to reach $u$ through $v$, and $\theta$ is simply the probability that $v$ has not yet transmitted to $u$.

Putting these arguments together, we define a \emph{test individual} $u$ to be a randomly chosen individual in the population which is prevented from transmitting to its partners.  Then $\theta(t)$ is the probability that a partner of $u$ has not yet transmitted to $u$.

When $u$ is taken to be a test individual, the definition of $p_I$ above becomes the probability that the partner $v$ is infected given that it has not yet transmitted to $u$.  In other words $p_I$ is the probability that $v$ is infected and has not transmitted given that it has not transmitted, and $\theta$ is the probability $v$ has not transmitted.  Then $p_I$ and $\theta$ are both probabilities, but they are conditional on different events.  This makes it difficult to work with these variables.  Effectively, $p_I$ is the ratio of two quantities, both of which are changing in time and so there is an implicit quotient rule hiding in the derivation of its derivative --- this is a challenge.  It is more natural to use $\phi_I=\theta p_I$ to be the probability that $v$ is infected and has not transmitted to $u$.  We similarly define $\phi_S$ to be the probability that $v$ is susceptible (and obviously has not transmitted to $u$) and $\phi_R$ to be the probability that $v$ is recovered and did not transmit to $u$.  With this $\theta = \phi_S+\phi_I+\phi_R$.

\subsection{Impact of preventing the test individual from transmitting} 
\label{sec:notrans}
One final concern may arise because modifying $u$ to prevent it from causing infection alters the dynamics of the epidemic.  Some individuals that would otherwise get infected may now remain susceptible, while others simply have their infection delayed.  We present two arguments for why this is not a concern.  For both of these arguments, we first note that once $u$ is infected, the time of its recovery is independent of any transmissions it causes.  So the modification of $u$ does not alter the probability that $u$ has a given status.

The first argument is that none of the effects of modifying $u$ are relevant.  Modifying $u$ does not affect its probability of being infected.  We have already seen that in the original epidemic (before $u$ is modified), the proportion of individuals in each state is equal to the probability $u$ is in each state.  We have a series of equivalent questions.  The first is, ``what proportions of the population are in each state in the original population?''  This is equivalent to our second question, ``what is the probability a randomly chosen individual $u$ is in each state in the original population?''  This is equivalent to our third question, ``what is the probability a randomly chosen individual $u$ is in each state if it is prevented from transmitting?''  At no point do we need to know anything in the modified population except the status of $u$, and preventing $u$ from transmitting in the modified population does not affect its status, it only affects the status of other individuals.  So the impact does not affect any quantities we calculate.

Our second argument is that in addition to not being relevant to the question we are asking, modifying $u$ has a negligible impact on the proportion infected in the population.  Although this is not needed for our argument here, it is relevant for derivation of final sizes~\cite{miller:final}.  To make this point, we use analogy to the ``price taker'' assumption of economics.  A firm is a price taker if it is too small to influence the price for its product.  Consequently, if all firms in a given market are price takers, we can determine how the actions of a given firm depends on the price, with the knowledge that its individual action does not affect the price.  Then we determine how the price depends on the collective actions of the entire market.  This will give a system of equations and we have a consistency relation which we can solve to find the strategies and resulting price.  We do not need to worry that an individual firm will have to change its strategy in response to the impact its individual strategy has on the price.  

When we assume that a stochastic process is behaving deterministically on some large aggregate scale, we are making a similar assumption.  In particular, for a disease spreading through a population, if we can assume that the aggregate dynamics are deterministic, then we are implicitly assuming that whether a particular individual is infected or not (and when that infection occurs) has no influence on the dynamics of the epidemic.  Not only does the individual's infection not have any measurable aggregate-scale impact, but also the infections traced back to that individual have no measurable aggregate-scale impact.  Thus when we consider a deterministic epidemic we can make an ``infection taker'' assumption: We can ignore the impact of modifying $u$ because it has no impact.\footnote{In fact, this explains why final sizes from epidemic simulations in smaller populations are often very similar to larger populations even if the dynamics are still highly stochastic: The timing of an individual's infection may have a significant impact on the aggregate number infected at any given time and therefore be important dynamically, even if it has little impact on the final size.} 

So for the purposes of determining the final proportion infected, we can calculate the probability an individual $u$ is infected given the amount of transmission that happens, but ignoring its impact on transmission.  Then we calculate the amount of transmission that will happen given the proportion of the population that is infected.  This leads to a consistency relation in which we know the proportion infected as a function of the proportion infected.

\section{Model hierarchy}
In this appendix we show that under reasonable assumptions, the models
presented in this paper are in fact equivalent.  We have three subtly
different closure approximations making slightly different assumptions
about the independence of partners.  Depending on which assumptions
hold, different models result, but all ultimately become identical in
appropriate limits.  We will show that by making appropriate
assumptions, we can derive some of the models from others.

\subsection{An example}
In several cases, the technique we use is a careful application of integrating
factors.  We demonstrate this technique with a different physical
problem for which most people's intuition is stronger.  

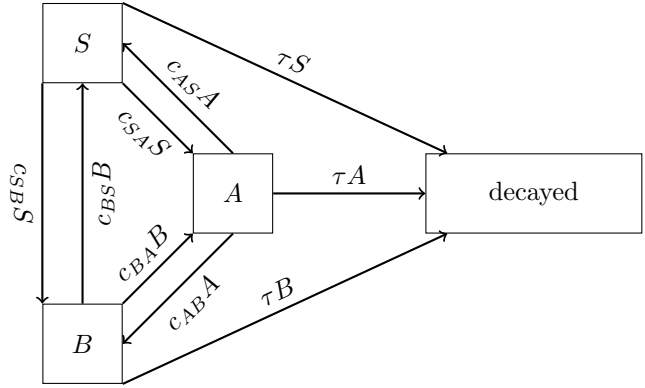
\begin{figure}
\begin{center}
\begin{tikzpicture}
\node [narrow] at (0,0) (A) {$A$};
\node [narrow] at (-2,2) (S) {$S$};
\node [narrow] at (-2,-2) (B) {$B$};
\node [wide] at (4,0) (decayed) {decayed};

\path [->, thick,sloped, above] (A.90) edge node {$c_{AS}A$} (S.0);
\path [->, thick,sloped, below] (A.270) edge node {$c_{AB}A$} (B.0);
\path [->, thick,sloped, above] (B) edge node {$c_{BA}B$} (A);
\path [->, thick,sloped, below] (B) edge node {$c_{BS}B$} (S);
\path [->, thick,sloped, below] (S) edge node {$c_{SA}S$} (A);
\path [->, thick,sloped,  below] (S.225) edge node {$c_{SB}S$} (B.135);
\path [->, thick, sloped, above] (A) edge node {$\tau A$} (decayed);
\path [->, thick, sloped, above] (B.315) edge node {$\tau B$} (decayed);
\path [->, thick, sloped, above] (S.45) edge node {$\tau S$} (decayed);
\end{tikzpicture}
\end{center}
\caption{\textbf{The flow diagram for the flux of a radioactive isotope
  between soil, biomass and air}.  The isotope decays at rate $\tau$.
  This model provides a useful example of a technique that will be
  used in later derivations.}
\label{fig:isotope}
\end{figure}

Let us assume there is a single release of a radioactive isotope into the environment.  The isotope may be in the air ($A$), in soil ($S$), or in biomass ($B$).  It decays in time with rate $\tau$ independently of where it is.  Assume the fluxes between the compartments are as in figure~\ref{fig:isotope}.

Then the equations are 
\begin{align*}
\dot{A} &= - \tau A + c_{SA}S - c_{AS}A + c_{BA}B-c_{AB}A \\
\dot{S} &=  - \tau S  - c_{SA} S + c_{AS} A - c_{SB}S - c_{BS} B \\
\dot{B} &= - \tau B  - c_{BA} B + c_{AB} A -c_{SB}S + c_{BS}B \, .
\end{align*} 

\begin{figure}
\begin{center}
\begin{tikzpicture}
\node [narrow] at (0,0) (A) {$a$};
\node [narrow] at (-2,2) (S) {$s$};
\node [narrow] at (-2,-2) (B) {$b$};

\path [->, thick,sloped, above] (A.90) edge node {$c_{AS}a$} (S.0);
\path [->, thick,sloped, below] (A.270) edge node {$c_{AB}a$} (B.0);
\path [->, thick,sloped, above] (B) edge node {$c_{BA}b$} (A);
\path [->, thick,sloped, below] (B) edge node {$c_{BS}b$} (S);
\path [->, thick,sloped, below] (S) edge node {$c_{SA}s$} (A);
\path [->, thick,sloped,  below] (S.225) edge node {$c_{SB}s$} (B.135);
\end{tikzpicture}
\end{center}
\caption{\textbf{The flow diagram for the flux of a radioactive isotope
  between soil, biomass and air}.  When we focus on a test particle which does not decay, the number of fluxes and compartments to consider is reduced.  This can be achieved mathematically through an integrating factor.}
\label{fig:isotope_nodecay}
\end{figure}
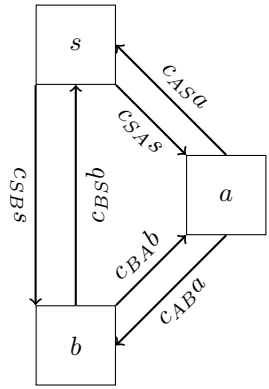

However, if we define $a$, $s$, and $b$ to be the probability that a test atom which does not decay is in each compartment, then the decayed class disappears.   We get the new flow diagram shown in figure~\ref{fig:isotope_nodecay}.  The new equations are \begin{align*}
\dot{a} &=  c_{SA}s - c_{AS}a + c_{BA}b-c_{AB}a \\
\dot{s} &=  - c_{SA} s + c_{AS} a - c_{SB}s - c_{BS} b \\
\dot{b} &=  - c_{BA} b + c_{AB} a -c_{SB}s + c_{BS}b \, .
\end{align*}
Physically this change of variables is fairly obvious.  We are
calculating the probability an isotope is in a given compartment
conditional on it having not yet decayed.

Mathematically we can get the new system of equations from the original through an integrating factor of $\alpha e^{\tau t}$.  We set $a = A \alpha e^{\tau t}$, \ $b=B \alpha e^{\tau t}$, and $c=C \alpha e^{\tau
  t}$ with $\alpha$ chosen so that the initial amounts sum to $1$.  If we multiply
the $\dot{A}$ equation by $e^{\tau t}$, we end up with
$\diff{}{t}(Ae^{\tau t}) = (c_{SA}S - c_{AS}A + c_{BA}B-c_{AB}A)
e^{\tau t}$.  Using the variable changes, we immediately arrive at the
$\dot{a}$ equation.  The other equations transform similarly.  So
using an integrating factor to eliminate the decay term is equivalent
to transforming into variables that measure the probability an 
undecayed isotope is in each compartment.

In general for other systems, so long as all compartments have an identical decay rate and the terms in the equations are homogeneous of order $1$, then it is possible to use an integrating factor in this way to define a change of variables that eliminates the decay term.  This will be a key step in deriving the EBCM approach from the other models.  Here the decay rate corresponding to infection of an individual from outside the pair plays this role.  

\subsection{Simplifications of basic pairwise model}
We begin by showing that the basic pairwise model can be reduced to the compact pairwise model, and that in turn, this is equivalent to the EBCM model.

\subsubsection{Deriving compact pairwise model from basic pairwise model}
We presented two pairwise models.  In both, we assumed the triples closure: Nothing we know about one partner of a susceptible individual $u$ gives any information about another partner of $u$.  We showed that the first reduces to
the second if we assume that given susceptible $u$ nothing we know about its degree gives any information about whether its partner $v$ is infected or susceptible.  Mathematically, this states that $\I_k=[S_kI]/k[S_k]$ and
$\Sa_k= [S_k S]/k[S_k]$ are independent of $k$.   The combination of these two assumptions gives us the pairs closure.  So under the pairs closure, we expect the compact pairwise model to hold.

Our derivation of the EBCM model was based on the pairs closure.  So we expect some change of variables to show that it is equivalent to the compact pairwise model.

It was previously noted~\cite{house:insights} that if we make the generic assumption that $[A_kB] = [AB] k[A_k]/\sum_l l [A_l]$ where $[A_kB]$ represents the number of partnerships between individuals of status $A$ having $k$ partners and individuals of status $B$, then a pairwise approach can be used to derive an early version of the EBCM model~\cite{volz:cts_time}.  In general, we expect this assumption to fail if $A$ is either $I$ or $R$.  However, in the particular case where status $A$ is susceptible, the assumption is consistent: Regardless of the degree of an individual, it has no impact on the status of its neighbors so long as it remains susceptible.  We do not need the general form of the closure for our derivation, just the particular case with $A=S$.

In the derivation of the compact pairwise model, we claimed that $\I$, the probability a partner of a susceptible individual $u$ is infected, is independent of $k$.  This follows from the pairs closure, but we did not prove that if we start with the basic pairwise model and assume this probability is independent of $k$, then it remains independent of $k$ at all later times.  To address this, we turn to $\I_k = [S_kI]/k[S_k]$ and $\Sa_k = [S_k S]/k[S_k]$.  We will take the derivative of $\I_k$, and show that if these are is initially $k$-independent, then its derivative is $k$-independent.  We have
 \begin{align*}
\dot{\I}_k &= \frac{\dot{[S_kI]}}{k[S_k]} - \frac{[S_kI] k
  \dot{[S_k]}}{k^2[S_k]^2}\\
&= \frac{\sum_{k'} \dot{[S_kI_{k'}]}}{k[S_k]} - \I_k
  \frac{k\dot{[S_k]}}{k[S_k]}\\
&= \frac{\sum_{k'} \beta
\frac{k'-1}{k'}\frac{[S_kS_{k'}][S_{k'}I]}{[S_{k'}]} - \gamma
[S_kI_{k'}] - \beta\frac{k-1}{k}\frac{[IS_k][S_kI_{k'}]}{[S_k]} -
\beta [S_kI_{k'}]}{k[S_k]} - \I_k \frac{k(-\beta[S_kI])}{k[S_k]}\\
&= \frac{-(\gamma+\beta - \beta(k-1)\I_k)[S_kI] + \sum_{k'} \beta
(k'-1)[S_kS_{k'}]\I_{k'}}{k[S_k]} - \I_k \frac{k(-\beta[S_kI])}{k[S_k]}\\
&= -(\gamma+\beta)\I_k - \beta(k-1)\I_k^2 + \frac{\sum_{k'} \beta
(k'-1)[S_kS_{k'}]\I_{k'}}{k[S_k]} + \beta k\I_k^2\\
&= -(\gamma+\beta)\I_k - \beta(k-1)\I_k^2 + \frac{\beta [S_kS]\sum_{k'}
 k'(k'-1)[S_{k'}]\I_{k'}}{k[S_k]\sum_{k''}k''[S_{k''}]} + \beta k\I_k^2\\
&= -(\gamma+\beta)\I_k - \beta(k-1)\I_k^2 + \frac{[S_kS]}{k[S_k]}\frac{\beta\sum_{k'}
k'(k'-1)[S_{k'}]\I_{k'}}{\sum_{k''}k''[S_{k''}]} + \beta k\I_k^2\\
&= -(\gamma+\beta)\I_k - \beta(k-1)\I_k^2 + \Sa_k\frac{\beta\sum_{k'}
k'(k'-1)[S_{k'}]\I_{k'}}{\sum_{k''}k''[S_{k''}]} + \beta k\I_k^2\\
&= -(\gamma+\beta)\I_k + \beta\I_k^2 + \Sa_k\frac{\beta\sum_{k'}
k'(k'-1)[S_{k'}]\I_{k'}}{\sum_{k''}k''[S_{k''}]} \, .
\end{align*}
So we see that if $\I_k$ and $\Sa_k$ are independent of $k$ at a given
time, then the derivative of $\I_k$ is also independent of $k$.  A
similar calculation shows that the derivative of $\Sa_k$ is
independent of $k$.  Thus we conclude that if at any time all $\I_k$ and all
$\Sa_k$ are independent of $k$, they remain so for future time.

This combined with the work in the main text shows that if $\I_k$ and $\Sa_k$ are initially $k$-independent (equivalently, the pairs closure holds), then the basic pairwise model reduces to the compact pairwise model.  We are now ready to derive the EBCM equations from the compact pairwise model.  

\subsubsection{Deriving EBCM model from compact pairwise model.}
We now derive the EBCM model from the compact pairwise model.   We later derive the compact pairwise model from the EBCM model. 

We begin our derivation with the observation that (for all $k$)
$\dot{[S_k]} = -\beta k \I [S_k]$.  So $[S_k](t) = [S_k](0) e^{-\beta
  k \int_0^t \I(\tau) \, \mathrm{d}\tau}$.  We define $\theta(t) =
e^{-\beta \int_0^t \I(\tau) \, \mathrm{d}\tau}$.  Then $[S_k](t) =
[S_k](0) \theta(t)^k$.  If we define $\psi(\theta) = \sum_k \frac{[S_k](0)}{N}
\theta^k$, then $S(t) = N\psi(\theta)$ and $\kex =
\theta\psi''(\theta)/\psi'(\theta)$.  We have $\dot{\theta} = -\beta
\I \theta$.

We return to the equations
\begin{align*}
\dot{[SS]} &= -2\beta  \kex \I [SS]\\
\dot{[SI]} &= \beta \kex \I [SS] - (\beta + \gamma) [SI] - \beta \kex \I [SI]\\
\dot{[SR]} &= \gamma [SI] - \beta \kex \I [SR] \, .
\end{align*}
We observe that each of these equations has a term which represents infection of the first individual in the partnership from a source outside the partnership (with the same coefficient, $\beta \kex \I$, each time).  So we anticipate that an integrating factor could eliminate this term.  In the EBCM approach, the test individual is prevented from causing infection.  This allows us to ignore infections to the test individual from individuals other than the partner along a given partnership.  Mathematically, this shows up as ignoring infections from outside the partnership of the first individual in the partnership.

We define $F(t)$ such
that $F'(t) = \beta \kex \I $.  There is an arbitrary constant in $F$
which we will choose later.  Then multiplying through by $e^{F(t)}$ we
find
\begin{align*}
\diff{}{t} ([SS] e^{F(t)}) &= - \beta \kex \I [SS] e^{F(t)}\\
\diff{}{t} ([SI] e^{F(t)}) &=  \beta \kex
\I[SS]e^{F(t)}-(\beta+\gamma)[SI]e^{F(t)}  \\
\diff{}{t} ([SR] e^{F(t)}) &= \gamma [SI] e^{F(t)} \, .
\end{align*}
We define $\phi_S = [SS]e^{F(t)}$, \ $\phi_I = [SI] e^{F(t)}$, and
$\phi_R = [SR] e^{F(t)}$.  Substituting these in and using $\kex = \theta \psi''(\theta)/\psi'(\theta)$, the equations become
\begin{align*}
\dot{\phi}_S &= - \beta \kex \I \phi_S = -\dot{\theta} \frac{\psi''(\theta)}{\psi'(\theta)}\phi_S\\
\dot{\phi}_I &= \beta \kex \I \phi_S - (\beta+\gamma) \phi_I =
-\dot{\phi}_S - (\beta+\gamma) \phi_I\\
\dot{\phi}_R &= \gamma \phi_I \, .
\end{align*}
By directly substituting in, we can check that $\phi_S = \frac{\psi'(\theta)}{\psi'(1)} \phi_S(0)$.

We choose the arbitrary constant in $F(t)$ such that
$\phi_S(0)+\phi_I(0)+\phi_R(0)=1$.  We note that the equations suggest
that there is some quantity being moved between compartments.  We have
$\diff{}{t} \phi_S+\phi_I+\phi_R = - \beta \phi_I$.  If we introduce
compartments, we have flux from $\phi_S$ to $\phi_I$, from $\phi_I$ to
$\phi_R$, and from $\phi_I$ to an as-yet-undefined compartment (which
will become $1-\theta$).

Since $\I = [SI]/([SS]+[SI]+[SR]) = \phi_I/(\phi_S+\phi_I+\phi_R)$,
we conclude that 
\begin{align*}
\diff{\theta}{t} &= -\beta \I \theta\\
& = - \beta \phi_I
\frac{\theta}{\phi_S+\phi_I+\phi_R}\\
&= \frac{\diff{\phi_S+\phi_I+\phi_R}{t} }{\phi_S+\phi_I+\phi_R} \theta \, .
\end{align*}
Since our initial condition has $\theta(0)=1 =
\phi_S(0)+\phi_I(0)+\phi_R(0)$, direct substitution into this equation
shows that $\theta = \phi_S+\phi_I+\phi_R$ is the solution.  So
$\dot{\theta} = -\beta \phi_I$.

Because $\dot{\phi}_R = -\gamma\dot{\theta}/\beta$, we can show that
$\phi_R = \phi_R(0) + \frac{\gamma}{\beta} (1-\theta)$.  
Using our equations for $\dot{\theta}$, our solutions for $\phi_S$ and
$\phi_R$, and our relation $\theta = \phi_S+\phi_I+\phi_R$, we
finally arrive at the EBCM
equations~\eqref{eqn:EBCMalpha}--\eqref{eqn:EBCMomega}.

\paragraph{Deriving compact pairwise model from EBCM model}

Our goal is to show that using the EBCM model we can use a change of variables to arrive at the compact pairwise equations.  Guided by the physical interpretation of our variables, our substitutions are:
\begin{align*}
[S]&=S\\
[I] &= I\\
[R] &= R\\
[S_k] &= S_k(0) \theta^k\\
[SS] &= \psi'(\theta)\phi_S\\
[SI] &= \psi'(\theta)\phi_I\\
\I &= \frac{\phi_I}{\theta}\\
\kex &= \theta \frac{\psi''(\theta)}{\psi'(\theta)} \, .
\end{align*}

Under these assumptions, we have
\[
\dot{[S_k]} = S_k(0) k \theta^{k-1} \dot{\theta} = - \beta S_k(0) k \theta^{k-1} \phi_I = - \beta k S_k(0) \theta^k \I = - \beta k [S_k] 
\]
which is the compact pairwise model equation for $\dot{[S_k]}$.  So the equation for $\dot{[S_k]}$ is as expected.

Since $[S] = S$ and $S = \sum_k S_k(0)\theta^k$, the equation for $[S]$ is $[S] = \sum_k S_k(0)\theta^k = \sum_k [S_k]$ as expected.  The equations for $[I]$ and $[R]$ are immediately as anticipated.

We now look at $[SS] = \psi'(\theta)\phi_S$.  We differentiate this to get
\begin{align*}
\dot{[SS]} &= \dot{\theta}\psi''(\theta)\phi_S + \psi'(\theta)\dot{\phi}_S\\
&= -\beta \phi_I \psi''(\theta)\phi_S + \psi'(\theta) \diff{}{t} \frac{\phi_S(0) \psi'(\theta)}{\psi'(1)}\\
&= -\beta \frac{\phi_I}{\theta} \theta\psi''(\theta)\phi_S + \psi'(\theta) \frac{\phi_S(0) \dot{\theta}\psi''(\theta)}{\psi'(1)}\\
&= -\beta \I \frac{\theta\psi''(\theta)}{\psi'(\theta)}\psi'(\theta)\phi_S -\beta\phi_I \psi'(\theta) \frac{\phi_S(0) \psi''(\theta)}{\psi'(1)}\\
&= -\beta \I \kex[SS] -\beta\frac{\phi_I}{\theta} \frac{\psi'(\theta)\phi_S(0) }{\psi'(1)}\theta\psi''(\theta)\\
&= -\beta \I \kex[SS] -\beta\I\phi_S \psi'(\theta)\frac{\theta\psi''(\theta)}{\psi'(\theta)}\\
&= -\beta \I \kex[SS] -\beta\I [SS]\kex\\
&= - 2\beta \I \kex [SS] \, .
\end{align*}
Which is the compact pairwise equation for $\dot{[SS]}$.

We next look at $[SI] = \psi'(\theta)\phi_I$.
We differentiate it to get
\begin{align*}
\dot{[SI]} &= \diff{}{t} [ \psi'(\theta)\phi_I]\\
&= \dot{\theta} \psi''(\theta)\phi_I + \psi'(\theta) \dot{\phi}_I\\
&= \dot{\theta}\psi''(\theta)\phi_I + \psi'(\theta) \diff{}{t} (\theta-\phi_S-\phi_R)\\
&= \dot{\theta}\psi''(\theta)\phi_I + \psi'(\theta) (\dot{\theta} - \dot{\phi}_S - \dot{\phi}_R)\\
&= \dot{\theta}\psi''(\theta)\phi_I + \psi'(\theta) (-\beta\phi_I - \dot{\phi}_S - \gamma\phi_I)\\
&= \dot{\theta}\psi''(\theta)\phi_I - \psi'(\theta) \dot{\phi}_S - (\beta+\gamma)[SI]\\
&= \dot{\theta} \frac{\psi'(\theta)\kex}{\theta} \phi_I- \psi'(\theta) \dot{\phi}_S - (\beta+\gamma)[SI]\\
&= (-\beta \phi_I) \frac{[SI] \kex}{\theta} - \psi'(\theta) \dot{\phi}_S - (\beta+\gamma)[SI]\\
&= -\beta \I [SI] \kex - \psi'(\theta) \dot{\phi}_S - (\beta+\gamma)[SI]\\
&= -\beta \I [SI] \kex - (\beta+\gamma)[SI] - \psi'(\theta) \dot{\phi}_S  \, .
\end{align*}
The term $\psi'(\theta)\dot{\phi}_S$ was shown above to be $-\beta \I \kex [SS]$.  So this becomes
\[
\dot{[SI]} = -\beta \I [SI] \kex - (\beta+\gamma)[SI]+\beta \I \kex [SS] 
\]
which is the compact pairwise equation for $\dot{[SI]}$.

 We now look at $\I$. We have assumed $\I = \phi_I/\theta$.  This becomes
\begin{align*}
\I &= \frac{\phi_I}{\theta}\\
&= \frac{[SI]}{\psi'(\theta)\theta}\\
&= \frac{[SI]}{\theta \sum_kS_k(0) k \theta^{k-1}}\\
&= \frac{[SI]}{\sum_kS_k(0) k \theta^k}\\
&= \frac{[SI]}{\sum_kk[S_k]}   
\end{align*}
which is the compact pairwise equation for $\I$.

We are finally left with $\kex$.  We have
\begin{align*}
\kex &= \theta \frac{\psi''(\theta)}{\psi'(\theta)}\\
&= \frac{\theta \sum_k k(k-1)S_k(0) \theta^{k-2}}{\sum_k k S_k(0) \theta^{k-1}}\\
&= \frac{\theta}{\theta}\frac{\theta \sum_k k(k-1)S_k(0) \theta^{k-2}}{\sum_k k S_k(0) \theta^{k-1}}\\
&= \frac{ \sum_k k(k-1)S_k(0) \theta^{k}}{\sum_k k S_k(0) \theta^{k}}\\
&= \frac{\sum_k k(k-1) [S_k]}{\sum_k k [S_k]}
\end{align*}
which is the compact pairwise equation for $\kex$.  

Thus we have derived all of the equations for the compact pairwise model from the EBCM model.  Since each model can be derived from the other, the models are equivalent.

\subsection{Simplification of the effective degree models}

Because of the order that different parts of this paper were developed, we find that rather than deriving the reduced effective degree model from the basic effective degree model, it is more practical to derive the basic pairwise model from the basic effective degree by adding the pairs closure.  Then we show that the reduced effective degree model is equivalent to the EBCM model, which we have already shown is equivalent to the basic effective degree model.  This suffices to prove that the reduced effective degree model can be derived from the basic effective degree model using the pairs closure.  For completeness, we later sketch the derivation of the reduced effective degree model from the basic effective degree model.

It should be noted that lthough we can recover the compact pairwise model from both the basic effective degree model and the basic pairwise model using the pair closure, we cannot recover the basic pairwise model from the basic effective degree model (or \emph{vice versa}).  The two basic models have slightly different underlying assumptions (the effective degree model makes the ``star closure'' while the basic pairwise model makes the ``triples closure''.  Neither is a special case of the other.

\subsubsection{Deriving compact pairwise model from basic effective degree model}
The effective degree model of~\cite{lindquist} can be used to derive a pairwise model closely related to the compact pairwise model we used.  Once the appropriate additional closure assumption is made, it becomes the compact pairwise model.  

To derive the compact pairwise model starting from the basic effective degree model, we begin by defining 
\begin{align*}
[ss] &= \sum_{s,i} sx_{s,i}\\
[si] &= \sum_{s,i}ix_{s,i} \, .
\end{align*}
These will represent the number of susceptible-susceptible partnerships
(counted twice, once with each partner as the first individual) and
the number of susceptible-infected partnerships.

We define 
\begin{align*}
[ssi] &= \sum_{s,i} s i x_{s,i}\\
[isi] &= \sum_{s,i} i (i-1) x_{s,i} \, .
\end{align*}
These will represent the number of triples of the
corresponding types.  We define
\begin{align*}
\hat{\xi} &= \frac{\sum_{s,i} is x_{s,i}}{ \sum_{s,i} s x_{s,i}} \\
&= \frac{[ssi]}{[ss]}
\end{align*}
and
\begin{align*}
\hat{\zeta} &= \frac{\sum_{s,i} i^2 x_{s,i}}{\sum_{s,i}i x_{s,i}}\\
&= \frac{\sum_{s,i} i(i-1) x_{s,i} + ix_{s,i}}{\sum_{s,i}i x_{s,i}}\\
&= \frac{[isi]}{[si]} + 1 \, .
\end{align*}
These correspond to the $\xi$ and $\zeta$ of the compact effective degree model.

We have 
\begin{align*}
\dot{[ss]} &= \sum_{s,i} s \dot{x}_{s,i}\\
&= \sum_{s,i} s(-\beta i x_{s,i}) + \sum_{s,i} s \gamma (i+1)x_{s,i+1} - \sum_{s,i} s
\gamma i x_{s,i} + \beta \xi\sum_{s,i} s [(s+1) x_{s+1,i-1} - s x_{s,i}]\\
&=-\beta \sum_{s,i} si x_{s,i} + \beta \xi\sum_{s,i} s [(s+1) x_{s+1,i-1} - s x_{s,i}]\\
&= - \beta \sum_{s,i} si x_{s,i}+ \beta \xi \sum_{s,i} (s+1)^2 x_{s+1,i-1} - \beta \xi \sum_{s,i} (s+1)
x_{s+1,i-1} - \beta \xi \sum_{s,i} s^2 x_{s,i}\\
&= - \beta \sum_{s,i} si x_{s,i} - \beta \xi \sum_{s,i} (s+1) x_{s+1,i-1}\\
&= - \beta \xi \sum_{s,i} s x_{s,i} - \beta \xi [ss]\\
&= - 2 \beta \xi [ss]\\
&= - 2 \beta [ssi] \, .
\end{align*}
Going from the third to the fourth line, we used the fact that $\sum_{s,i}
s(i+1)x_{s,i+1} = \sum_{s,i} si x_{s,i}$, and going from the fourth to the fifth line we used $\sum_{s,i} (s+1)^2 x_{s+1,i-1} =
\sum_{s,i} s^2 x_{s,i}$.
We further have
\begin{align*}
\dot{[si]} &= \sum_{s,i} i \dot{x}_{s,i}\\
&= \sum_{s,i} i (-\beta i x_{s,i}) + \gamma \sum_{s,i} i(i+1)x_{s,i+1} - \gamma
\sum_{s,i} i^2 x_{s,i} + \beta \xi \sum_{s,i} i[(s+1)x_{s+1,i-1} - s x_{s,i}]\\
&= - \beta \sum_{s,i} i^2 x_{s,i} + \gamma \sum_{s,i} (i+1)^2 x_{s,i+1} - \gamma
\sum_{s,i} i^2 x_{s,i} - \gamma \sum_{s,i} (i+1) x_{s,i+1} + \beta \xi \sum_{s,i} i[(s+1)x_{s+1,i-1} - s x_{s,i}]\\
&= - \gamma [si] - \beta \sum_{s,i} i^2 x_{s,i}+ + \beta \xi \sum_{s,i} (i-1)
(s+1) x_{s+1,i-1} + \beta \xi \sum_{s,i} (s+1) x_{s+1,i-1} - \beta \xi \sum_{s,i}
si x_{s,i} \\
&= - \gamma [si] - \beta \sum_{s,i} i^2 x_{s,i}+ \beta \xi \sum_{s,i} (s+1)
x_{s+1,i-1}\\
&= - \beta \zeta [si] - \gamma [si] + \beta \xi [ss]\\
&= - \beta [isi] - \beta [si] - \gamma[si] + \beta [ssi] \, .
\end{align*}
These are the equations of the global (unclosed) pairwise model at the level of
pairs without using either the triples or the pairs closure.

Rather than using the triples closure to simplify the triples terms in these equations, we use the star closure, or more precisely, we use the basic effective degree model to simplify the triples terms.  We can derive equations for the rate
of change of $[isi] = \sum_{s,i} i(i-1) x_{s,i}$ and the rate of change of
$[ssi] = \sum_{s,i} si x_{s,i}$ simply by using the derivative of $x_{s,i}$, equation~\eqref{eqn:lindquistalpha}.  When we use this, we are introducing the star closure to the global (unclosed) pairwise model.  Although in general the rate of change of $[isi]$ and $[ssi]$ should depend on groupings of four individuals either as 3 individuals connected to a central individual or 4 individuals in a path, when we assume the derivative of $x_{s,i}$ from equation~\eqref{eqn:lindquistalpha}, we are assuming that we can safely average out the case of 4 individuals in a path.  This is the simplification of the star closure.  So we can express the basic effective degree model in terms of pairs and triples rather than effective degree.

We now add the pairs closure to this model.  Let us choose a random $ss$ partnership and label one individual $u$ and the other $v$.  Because the partnership is randomly chosen, we do not know the effective degree of $u$.  Note that $u$ is chosen with probability proportional to the number of $ss$ partnerships it is in.  The expected number of additional partners of $u$ is $\kex = \sum_{s,i} s(i+s-1) x_{s,i} / \sum_{s,i} s x_{s,i}$.  The probability that a partner $w \neq v$ is infected is $\I = \sum_{s,i} si x_{s,i} \sum_{s,i} s(i+s-1) x_{s,i}$ (the numerator gives the number of ways to choose $v$, $u$, and $w$ such that $v$ and $u$ are susceptible and $w$ infected while the denominator gives the number of ways to choose $v$ and $u$ to be susceptible and $w$ to be either susceptible or infected).  Thus the number of $ssi$ triples is $[ssi]=[ss]\kex\I$.  If we repeat this argument with an $si$ partnership having $u$ susceptible and $v$ infected, we will get a new expression for the probability $w$ is infected: $\I = \sum_{s,i} i(i-1) x_{s,i}/\sum_{s,i} i(i+s-1) x_{s,i}$.  However, the pairs closure means both values of $\I$ are the same and both vales of $\kex$ are the same.  So we find $[isi] = [is] \kex \I$.  Plugging this in we have
\begin{align*}
\dot{[ss]} &= -2\beta [ss] \kex \I\\
\dot{[si]} &= -\beta [is] \kex \I - (\beta+\gamma)[si] + \beta [ss] \kex \I \, .
\end{align*}
These are the equations we need for the pairs in the compact pairwise model,
equations~\eqref{eqn:compactalpha}--\eqref{eqn:compactomega}.  The remaining equations follow easily.

\subsubsection{Equivalence of the reduced effective degree model and the EBCM model}

\paragraph{Transforming reduced effective degree model to EBCM model}

We begin with the reduced effective degree model, equations~\eqref{eqn:ball_improved_alpha}--\eqref{eqn:ball_improved_omega}.  
We will look at the equation for $\dot{\nu}$ and an equation for $\diff{}{t} \sum_j j x_j$.
Our goal is to derive the EBCM equations.  We anticipate that $\nu$ should be proportional to $\phi_I$ and $\sum_j j x_j$ should be proportional to $\phi_S+\phi_I$.  When we consider an individual $u$ and a partner $v$ in the EBCM approach, we never have to consider the effect of $u$ being infected by a partner other than $v$.  Guided by this, we expect an integrating factor to allow us to throw out terms that represent infection coming from outside the partnership.  We begin with $\sum_j j x_j$.  This sums all of the partnerships from the perspective of each susceptible individual.  We look for a term that corresponds to eliminating a partnership from consideration because the focal individual is infected along a different partnership.
\begin{align*}
\diff{}{t} \sum_j j x_j &= \sum_j j \dot{x}_j\\
&= \gamma \I \sum_j\big((j+1)jx_{j+1} - j^2 x_j \big) - \beta \I \sum_j j^2 x_j\\
&= \gamma \I \sum_j\big((j+1)jx_{j+1} - j^2 x_j \big) - \beta \I \sum_j \big(j(j-1) x_j + j x_j\big)\\
&= \gamma \I \sum_j\big((j+1)jx_{j+1} - j^2 x_j \big) - \beta \I \sum_j j(j-1) x_j - \beta \I \sum_j j x_j\big)\\
&= \gamma \I \sum_j\big((j+1)jx_{j+1} - j^2 x_j \big) - \beta \I \frac{\sum_j j(j-1) x_j}{\sum_j j x_j} \sum_j j x_j - \beta \I \sum_j j x_j \, .
\end{align*}
The term $- \beta \I \frac{\sum_j j(j-1) x_j}{\sum_j j x_j} \sum_j j x_j$ is exactly the term of interest.  To remove it, we set $F'(t) = \beta \I \frac{\sum_j j(j-1) x_j}{\sum_j j x_j} $ and multiply by $e^{F(t)}$ (there is an arbitrary constant of integration which we will set later).  We define $\phi_S+\phi_I = e^{F(t)}\sum_j j x_j$.
\begin{align*}
\diff{}{t} (\phi_S+\phi_I) &= \gamma \I \sum_j\big((j+1)jx_{j+1} - j^2 x_j \big)e^{F(t)} - \beta \I \sum_j j x_j e^{F(t)} \\
&= \gamma \I \left( \sum_j \big((j+1)^2-(j+1)\big)x_{j+1} - \sum_j j^2 x_j \right)e^{F(t)} - \beta \I (\phi_S+\phi_I)\\
&= -\gamma \I \sum_j (j+1)x_{j+1} e^{F(t)} - \beta \I (\phi_S+\phi_I)\\
&= -\gamma \I (\phi_S+\phi_I)- \beta \I (\phi_S+\phi_I) \, .
\end{align*}
We now move to $\nu$.  This counts the number of susceptible-infected partnerships.  We look for the term that corresponds to infection of the susceptible individual from outside the partnership.
\begin{align*}
\dot{\nu} &= -(\beta+\gamma)\nu + \beta \I(1-2\I) \sum_j j(j-1)x_j\\
&= -(\beta+\gamma)\nu + \beta\I (1-\I) \sum_j j (j-1) x_j - \beta \I^2 \sum_j j (j-1) x_j\\
&= -(\beta+\gamma)\nu + \beta\I (1-\I) \sum_j j (j-1) x_j - \beta \I  \frac{\sum_j j (j-1) x_j}{\sum_j j x_j} \nu \, .
\end{align*}
The final term is the term we want.  Not coincidentally, the coefficient is the same as we saw above for $\sum_j j x_j$.  We use the same integrating factor and define $\phi_I = \nu e^{f(t)}$. We have
\begin{align*}
\dot{\phi}_I &= -(\beta+\gamma) \nu e^{F(t)}+ \beta\I (1-\I) \sum_j j (j-1) x_je^{F(t)}\\
&= -(\beta+\gamma) \phi_I + \beta \I \left(1-\frac{\nu}{\sum_j j x_J}\right) \sum_j j (j-1) x_j e^{F(t)}\\
&= -(\beta+\gamma) \phi_I + \beta \I \left(\frac{\sum_j j x_j - \nu}{\sum_j j x_J} e^{F(t)}\right) \sum_j j (j-1) x_j\\
&= -(\beta+\gamma) \phi_I + \beta \I \phi_S \frac{\sum_j j (j-1) x_j}{\sum_j j x_j}\\
&= -(\beta+\gamma) \phi_I + \phi_S F'(t) \, .
\end{align*}
Note that it is now clear that $\I = \phi_I/(\phi_S+\phi_I)$.  So we can simplify somewhat.  Our expression for $diff{}{t} (\phi_S+\phi_I)$ becomes $-(\beta+\gamma)\phi_I$.  Using the equations
\begin{align*}
\diff{}{t} (\phi_S+\phi_I) &= -(\beta+\gamma)\phi_I\\
\diff{}{t} \phi_I &= -(\beta+\gamma)\phi_I +\phi_S F'(t) 
\end{align*}
we finally have
\begin{align*}
\dot{\phi}_S &= -\phi_S F'(t)\\
\dot{\phi}_I &= -(\beta+\gamma)\phi_I + \phi_S F'(t)\, .
\end{align*}

We now introduce two variables $\theta$ and $\phi_R$.  We take $\theta(0)=1$, \ $\dot{\theta} = -\beta \phi_I$, and $\phi_R = \theta - \phi_S- \phi_I$.  It follows that $\dot{\phi}_R = \gamma \phi_I$.  Note that as yet, $\phi_S$ and $\phi_I$ are only known up to a multiplicative constant (due to the constant of integration in $F$).  We can arbitrarily set $\phi_S(0)+\phi_I(0)=1$, so that $\phi_R(0)=0$.  In terms of the interpretation of the EBCM variables, this corresponds to saying that we can ignore individuals that are already recovered at $t=0$.  We can choose another constant in $(0,1)$ if we wish to account for the already recovered partners of random individuals.

We turn our attention to finding $x_j$ in terms of our new variables, which will allow us to determine $F'(t)$ in terms of our new variables.  Guided by our expectations, we assume $x_j = \sum_{k \geq j} x_k(0) \binom{k}{j}(\phi_S+\phi_I)^j\phi_R^{k-j}$.  At $t=0$, this is correct.  We now check that the evolution of $x_j$ is correctly captured by this equation.  We find
\begin{align*}
\dot{x}_j &= \sum_{k \geq j} x_k(0) \binom{k}{j}\diff{}{t} \big((\phi_S+\phi_I)^j\phi_R^{k-j}\big)\\
&= \sum_{k \geq j} x_k(0) \binom{k}{j} \big ( -j(\beta+\gamma)\phi_I(\phi_S+\phi_I)^{j-1}\phi_R^{k-j} + \gamma(k-j)\phi_I(\phi_S+\phi_I)^j\phi_R^{k-j-1}\big)\\
&= \frac{\phi_I}{\phi_S+\phi_I}\sum_{k \geq j} x_k(0) \binom{k}{j} \big ( -j(\beta+\gamma)(\phi_S+\phi_I)^{j}\phi_R^{k-j} + \gamma(k-j)(\phi_S+\phi_I)^{j+1}\phi_R^{k-j-1}\big)\\
&= \frac{\phi_I}{\phi_S+\phi_I}\sum_{k \geq j} x_k(0)  \left ( -\binom{k}{j}j(\beta+\gamma)(\phi_S+\phi_I)^{j}\phi_R^{k-j} + \binom{k}{j}(k-j)\gamma(\phi_S+\phi_I)^{j+1}\phi_R^{k-j-1}\right)\\
&= \I\big( -(\beta+\gamma)j x_j +\gamma x_{j+1} \big) \, .
\end{align*}
This is the equation for $\dot{x}_j$.  So our ansatz is correct.

We now define $\psi(\theta) = \sum_k x_k(0)
\theta^k$ so that 
\[
\sum_j x_j = \sum_k \sum_j x_k(0) \binom{k}{j}
\phi_R^{k-j}(\phi_S+\phi_I)^j = \sum_k x_k(0)(\phi_S+\phi_I+\phi_R)^k
= \psi(\theta) \, .
\]
We will use this to express $F'$ in terms of the EBCM variables.  We first derive two identities using the ``derivative trick'':
\begin{align*}
\sum_{j=0}^\infty \binom{k}{j} j a^j b^{k-j} &= a \sum_{j=0}^\infty \binom{k}{j} j a^{j-1} b^{k-j}\\
&= a\diff{}{a} \sum_{j=0}^\infty \binom{k}{j} a^j b^{k-j}\\
&= a \diff{}{a} (a+b)^k\\
&= k a(a+b)^{k-1} \, ,
\end{align*}
and
\begin{align*}
\sum_{j=0}^\infty \binom{k}{j} j(j-1) a^j b^{k-j} &= a^2 \sum_{j=0}^\infty \binom{k}{j} j(j-1) a^{j-2} b^{k-j}\\
&= a^2\diffm{2}{}{a} \sum_{j=0}^\infty \binom{k}{j} a^j b^{k-j}\\
&= a^2 \diffm{2}{}{a} (a+b)^k\\
&= k(k-1) a^2(a+b)^{k-2} \, .
\end{align*}
So 
\begin{align*}
F'(t) &= \beta \I \frac{\sum_j j(j-1) x_j}{\sum_j j x_j} \\
&= \beta \I \frac{\sum_k \sum_j j(j-1) x_k(0) \binom{k}{j} (\phi_S+\phi_I)^j\phi_R^{k-j}}{\sum_k \sum_j j x_k(0)\binom{k}{j} (\phi_S+\phi_I)^j\phi_R^{k-j}}\\
&= \beta \I \frac{(\phi_S+\phi_I)^2 \sum_k k(k-1)x_k(0) \theta^{k-2}}{(\phi_S+\phi_I)\sum_k k x_k(0) \theta^{k-1}}\\
&= \beta \I (\phi_S+\phi_I) \frac{\psi''(\theta)}{\psi'(\theta)}\\
&= \beta \phi_I \frac{\psi''(\theta)}{\psi'(\theta)} \, .
\end{align*}
So we now have
\[
\dot{\phi}_S = -\phi_S F'(t) = -\phi_S \beta \phi_I \frac{\psi''(\theta)}{\psi'(\theta)} \, .
\]
We make the observation that $\phi_S = \phi_S(0)\psi'(\theta)/\psi'(1)$ solves the ODE for $\phi_S$.  We have $\dot{\phi}_R = \gamma \phi_I$ and $\phi_I = \theta-\phi_S-\phi_R$.  This combined with $\dot{\theta}=-\beta\phi_I$ completes our system and gets us back to the EBCM equations.

So we have shown that starting from the reduced effective degree equations, we can derive the EBCM equations with $\phi_R(0)=0$.  The assumption that $\phi_R(0)= 0$ forced us to use $x_k(0)$ rather than $S_k(0)$ for our definition of $\psi$.  We could have used $S_k(0)$, but this would have required a different value of $\phi_R(0)$.  The derivation would have been more complicated, but not significantly altered.

\paragraph{EBCM to reduced effective degree}
We now go in the opposite direction, we derive the  reduced effective degree equations from the EBCM equations.  This is less strenuous.  We simply determine what the reduced effective degree variables should be in terms of the EBCM variables and then test that they satisfy the appropriate equations.  We can drop the assumption that $\phi_R(0)=0$.  We start with the EBCM equations and write
\[
x_{j}(t) = \sum_k S_k(0) \binom{k}{j} (\phi_S(t)+\phi_I(t))^j\phi_R(t)^{k-j}
 \]
[this is similar to the ansatz we used above for $x_j$, except we take $S_k(0)$ rather than $x_k(0)$].  We also anticipate that $\nu = \phi_I \psi'(\theta)$.  This is expected because $\psi'(\theta)$ gives the sum of degrees of susceptible individuals and $\phi_I$ gives the probability a partner of a susceptible individual is infected.

To test the equation for $x_j$ we follow the steps we did above for the ansatz of $x_j$.  These steps arrive at the expected ODE for $x_j$.

To test the equation for $\nu$, we first recall that $\phi_S=\phi_S(0) \psi'(\theta)/\psi'(1)$.  This will be used below.  We have
\begin{align*}
\dot{\nu} &= \diff{}{t} \phi_I \psi'(\theta)\\
&= \left( - (\beta+\gamma)\phi_I + \frac{\phi_S(0)}{\psi'(1)} \beta \phi_I \psi''(\theta)\right)\psi'(\theta) + \phi_I(-\beta\phi_I)\psi''(\theta)\\
&= -(\beta+\gamma)\phi_I\psi'(\theta) + \frac{\phi_S(0)}{\psi'(1)}\psi'(\theta)\beta\phi_I\psi''(\theta) - \beta \phi_I^2 \psi''(\theta)\\
&= -(\beta+\gamma)\nu + \phi_S \beta \phi_I \psi''(\theta) - \beta \phi_I^2\psi''(\theta)\\
&= -(\beta+\gamma)\nu + \beta \phi_I \psi''(\theta) (\phi_S-\phi_I)\\
&= -(\beta+\gamma)\nu + \beta \frac{\phi_I}{\phi_S+\phi_I}(\phi_S+\phi_I) \psi''(\theta) (\phi_S+\phi_I-2\phi_I)\\
&= -(\beta+\gamma)\nu + \beta \I (\phi_S+\phi_I)^2\psi''(\theta) (1-2\I)\\
&= -(\beta+\gamma)\nu + \beta \I (1-2\I)(\phi_S+\phi_I)^2\sum_k k(k-1)S_k(0)\theta^{k-2} \, ,
\end{align*}
Using $\theta=(\phi_S+\phi_I)+\phi_R$ and the derivative trick used above, we know $S_k(0)k(k-1)(\phi_S+\phi_I)^2\theta^{k-2}$ is $\sum_j \binom{k}{j} j(j-1)S_k(0) (\phi_S+\phi_I)^j\phi_R^{k-j}$.  So the final term above becomes
$\beta \I (1-2\I) \sum_J j(j-1) \sum_k S_k(0) (\phi_S+\phi_I)^j \phi_R^{k-j} = \beta \I (1-2\I) \sum_j j(j-1) x_j$.
Thus we have
\[
\dot{\nu} = -(\beta+\gamma)\nu + \beta \I(1-2\I) \sum_j j (j-1)x_j \, .
\]
So we can derive the reduced pairwise equations from the EBCM equations.  The two models are equivalent.

\subsubsection{Deriving reduced effective degree model from basic effective degree model}

We now address the direct relation between the basic effective degree model and the reduced effective degree model.  The two models use slightly different definitions of effective partnerships, but these deinitions coincide in the case of susceptible individuals.  This allows us to derive the simplified equations from the basic effective degree model.  We have already shown that it is possible to derive the reduced effective degree model from the basic effective degree model by a series of indirect steps deriving other models, so we only sketch this derivation.

We can define $x_{k_e} = \sum_{i+s=k_e} x_{s,i}$ where
$x_{s,i}$ is the variable from the basic effective degree model.  We find
\[
\dot{x}_{k_e} = \sum_{s+i=k_e} \gamma (i+1) x_{s,i+1} - \gamma i x_{s,i} - \beta i
x_{s,i} = \sum_{s+j=k_e+1} \gamma j x_{s,j} - \sum_{s+i={k_e}} \gamma i
x_{s,i} - \sum_{s+i={k_e}} \beta i x_{s,i} \, .
\]
We set $\I_{k_e} = \sum_{s+i={k_e}} i x_{s,i} /
\sum_{s+i={k_e}} {k_e}x_{s,i}$ and make the assumption that $\I_{k_e}$ is
in fact independent of ${k_e}$ (that is, the probability a partner of a
susceptible individual $u$ is infected is independent of the number of
effective partnerships $u$ has which follows from the pairs closure), then our equation reduces to
\[
\dot{x}_{k_e} = \gamma \I ({k_e}+1)x_{{k_e}+1} - \gamma k_e\I x_{k_e} - \beta k_e\I x_{k_e}
\]
which is equation~\eqref{eqn:ball_improved_alpha} with $k_e$ replacing $j$.

If we define $\nu = \sum_s\sum_i i x_{s,i}$ we get
\begin{align*}
\dot{\nu} &= \sum_s \sum_i i \dot{x}_{s,i}\\
&= \sum_s \sum_i \gamma i [(i+1)x_{s,i+1}-ix_{s,i}] - \beta i^2 x_{s,i} + \beta i \xi[ (s+1)x_{s+1,i-1}-sx_{s,i}] \, .
\end{align*}
After taking advantage of the fact that, for example $\sum_s \sum_i ix_{s,i} = \sum_s \sum_i (i+1)x_{s,i+1}$ and repeatedly reindexing som sums, and the fact that $\xi = \sum_s\sum_ six_{s,i}/\sum_s\sum_i s x_{s,i}$ we can convert this to
\[
\dot{\nu} = -(\gamma+\beta) \nu + \beta \sum_s \sum_i -i(i-1) x_{s,i} + \beta\sum_s\sum_isix_{s,i} \, .
\]
The sum $\sum_s\sum_i i(i-1) x_{s,i}$ counts the total number of (ordered) triples $(v,u,w)$ where $u$ has partnerships with both $v$ and $w$, $u$ is susceptible, and $v$ and $w$ are both infected.  We find this value through a different accounting based on the pairs closure: We count the total number of triples having a susceptible central individual.  The probability that both peripheral individuals are infected is $\I^2$.  We get $\I^2 \sum_{k_e} k_e(k_e-1)x_{k_e}$ where $x_{k_e}$ is as above.  Similarly, $\sum_s \sum_i six_{s,i}$ becomes $\I(1-\I)\sum_{k_e} k_e(k_e-1)x_{k_e}$.  Thus we finally have
\[
\dot{\nu} = -(\beta+\gamma)\nu + \beta\I(1-2\I) \sum_{k_e} k_e(k_e-1)x_{k_e} 
\]
which is the equation for $\nu$ from before.  

To complete this derivation, we would need to additionally show that $\I_{k_e}$ remains independent of $k_e$ if the initial conditions satisfy the pairs closure.  This will require expanding the derivative of $\I_{k_e}$ and showing that if initially these are independent of $k_e$, then the derivatives are independent of $k_e$.

\subsubsection{Deriving the compact effective degree model from the reduced effective degree model}

To derive the compact effective degree model, we start with $\nu = \I \sum_j jx_j$.  Anticipating that the two models to have the same value for $\I$, we note that in the compact effective degree model, $\lambda = \lambda \I + \I \sum jx_j$.  Solving this, we find $\lambda = \nu/(1-\I)$.  So starting from the reduced model equations, we define $\lambda = \nu/(1-\I)$.  Taking the derivative of this and using the same sort of changes done above will show that the derivative of $\lambda$ is indeed as found for the compact effective degree model.  

\providecommand{\noopsort}[1]{}

\end{document}